\documentclass[letterpaper,man,floatsintext,donotrepeattitle]{apa7}

\usepackage[american]{babel}
\usepackage[utf8]{inputenc}
\usepackage{amsmath}
\usepackage{graphicx}
\usepackage{xcolor, colortbl}
\usepackage{caption}
\usepackage{setspace}
\usepackage{threeparttable}
\usepackage{endnotes}
\usepackage{xspace}
\usepackage{afterpage}
\usepackage{mathtools}
\mathtoolsset{showonlyrefs}
\usepackage{soul}
\usepackage{bm}
\usepackage{dsfont}
\usepackage{amssymb}
\usepackage{lscape}
\usepackage{hyperref}
\hypersetup{
    colorlinks=true,
    allcolors=black
}
\usepackage[normalem]{ulem}
\usepackage{csquotes}
\usepackage{algorithm}
\usepackage[]{algpseudocode} 
\usepackage{amsthm}
\theoremstyle{remark}

\usepackage[style=apa,sortcites=true,sorting=nyt,backend=biber]{biblatex}
\DeclareLanguageMapping{american}{american-apa}
\usepackage{array}
\newcommand{\PreserveBackslash}[1]{\let\temp=\\#1\let\\=\temp}
\newcolumntype{C}[1]{>{\PreserveBackslash\centering}p{#1}}
\newcolumntype{L}[1]{>{\PreserveBackslash\raggedright}p{#1}}
\newcolumntype{R}[1]{>{\PreserveBackslash\raggedleft}p{#1}}

\usepackage{accents}

\newcommand{\indep}{\perp\!\!\!\perp}

\usepackage{arydshln}
\usepackage{ragged2e}
\usepackage{multirow}
\usepackage{subcaption}
\usepackage{amsmath,amssymb,xcolor}
\usepackage{pgf,tikz}
\usetikzlibrary{arrows,shapes.arrows,shapes.geometric,shapes.multipart, decorations.pathmorphing,positioning,shapes,decorations,arrows,calc,arrows.meta,fit,positioning}
\tikzset{
    -Latex,auto,node distance =1 cm and 1 cm,semithick ,
    state/.style ={ellipse, draw, minimum width = 0.7 cm},
    point/.style = {circle, draw, inner sep=0.04cm,fill,node contents={}},
    bidirected/.style={Latex-Latex,dashed},
    el/.style = {inner sep=2pt, align=left, sloped},
    styRectDef/.style = {rectangle, rounded corners, draw=black, inner xsep=6mm, inner sep=3mm}
}

\title{\large Causal Machine Learning for Multilevel Studies with Unmeasured Cluster-Level Covariates via Variational Autoencoders}

\shorttitle{VAE-R for Multilevel Studies}

\author{Youmi Suk and Yuxuan Li}

\authorsaffiliations{Teachers College Columbia University \break \texttt{\{ysuk{,} yl5655\}@tc.columbia.edu} \vspace*{-20pt}}

\abstract{Unmeasured confounders are a key obstacle to accurately estimating treatment effects in observational studies, and in multilevel data, they often arise at the cluster level. Flexible machine learning methods cannot directly resolve such unmeasured confounding. To address this, we propose the variational autoencoder-augmented R-learner (VAE-R), which leverages variational autoencoders (VAEs) to infer unmeasured cluster-level covariates from the within-cluster dependence of outcome and treatment residuals, and uses the resulting adjustments to estimate the conditional average treatment effect (CATE) and the average treatment effect (ATE). The VAE decoder is specified through an offset likelihood so that the latent variables capture only the cluster-level variation left unexplained by measured covariates. VAE-R accommodates unmeasured cluster-level confounders in observational studies and unmeasured cluster-level predictors of the outcome in multisite randomized trials. We also provide data-driven strategies for selecting the two hyperparameters of the VAE, the latent dimension size and the Kullback-Leibler weight. Our simulation study shows that VAE-R nearly eliminates the bias of the ordinary R-learner's CATE and ATE estimates in multilevel observational data and improves their precision in multisite randomized trials. Finally, we illustrate VAE-R with two datasets, one from a multilevel observational study and the other from a multisite randomized trial.}


\begin{document}

{
\setlength{\topskip}{0pt}
\vspace*{-120pt}  
\let\clearpage\relax
\let\newpage\relax
\maketitle
}

\vspace*{5pt} 

\setcounter{secnumdepth}{3}

\section{Introduction}

\subsection{Motivation}

For causal inference in observational studies, the assumption of ``no unmeasured confounding'' (also known as unconfoundedness or ignorability) is a fundamental requirement for methods based on matching or propensity score designs \parencite{imbens2015causal, rosenbaum2002observational}. Confounders are variables that affect both the treatment assignment and the outcome. However, it is often difficult to measure all confounders in practice. If some are unmeasured, they introduce bias into effect estimates and can severely undermine program evaluations. Historically, researchers have addressed this issue with four main strategies: (i) employing alternative quasi-experimental designs that do not require the unconfoundedness assumption, such as regression discontinuity, instrumental variables, and synthetic control designs \parencite[e.g.,][]{angrist1996identification, thistlethwaite1960regression}; (ii) adopting methods that adjust for unmeasured confounding using observed proxy variables \parencite[e.g.,][]{lee2021partially, suk2022robust, TT2024_Proximal}; (iii) conducting sensitivity analyses to determine whether a hypothesized unmeasured confounder alters the causal conclusion \parencite[e.g.,][]{rosenbaum2002observational}; and (iv) inferring unmeasured or latent confounders from observed data to adjust the causal analysis \parencite[e.g.,][]{morimura2025latent, louizos2017CEVAE}. This paper takes the fourth approach by learning latent cluster-level representations in multilevel studies from the dependence among residuals, using a well-established deep generative model known as the variational autoencoder \parencite[VAE;][]{kingma2014auto} 

Throughout, we use \emph{multilevel observational studies} and \emph{multisite randomized trials} to refer to studies in which individuals are nested within clusters (e.g., students within schools) and the treatment is assigned at the individual level; the two differ in whether treatment is randomized within clusters. In multilevel observational studies, researchers typically encounter two types of confounders: cluster-level confounders, which are shared by all individuals within a given cluster, and individual-level confounders, which vary across individuals. For example, suppose that a researcher investigates the effect of taking an advanced math course on students' math achievement using data from the High School Longitudinal Study of 2009 (HSLS), a nationally representative longitudinal study where students are nested within schools. Potential cluster-level confounders include school funding and school's emphasis on mathematics. 

In contrast, cluster-level covariates do not bias treatment effect estimates in multisite randomized trials, where treatment is randomly assigned within each cluster and the assignment mechanism is known to researchers by design. However, researchers are often interested in improving the precision of these estimates, for example by incorporating predictors that affect the outcome (i.e., prognostic variables) into the analysis \parencite{gagnon2023precise}. Consider evaluating the effect of a growth mindset intervention on academic achievement using data from a multisite randomized trial like the National Study of Learning Mindsets (NSLM). In this context, school-specific grading policies and instructional funding are potential predictors of the outcome. However, they are often unmeasured, and omitting them forgoes an opportunity to obtain more precise treatment effect estimates. Such predictors can improve precision in multilevel observational studies as well, but in that setting, their omission is a secondary concern because it does not induce confounding bias. 

If a researcher were able to measure all individual-level and cluster-level confounders in a multilevel observational study, they could use existing modern causal inference methods. In particular,  machine learning (ML) methods, such as causal forests \parencite{wager2018estimation} and double/debiased machine learning \parencite[DML;][]{chernozhukov2018doubleml}, have become increasingly popular due to their flexible learning and automatic fitting procedures. However, when key confounders are unmeasured, these estimators can be biased even if they can flexibly estimate the propensity score or the outcome model. For example, HSLS data do not fully capture  cluster-level confounders such as schools' instructional funding and their emphasis on math. In the presence of such unmeasured cluster-level confounders, standard ML methods for causal inference are likely to produce biased estimates of the target treatment effect. Similarly, while omitting cluster-level predictors does not cause bias, it yields less precise estimates of the treatment effect. 

\subsection{Prior Work and Our Contributions}

Initial approaches to addressing cluster-level unmeasured confounding relied on parametric fixed- or random-effects models for either the outcome or the treatment assignment \parencite{arpino2011specification, li2013propensity, firebaugh2013fixed,li2020impact}. However, these early methods are vulnerable to bias from mis-specifying either the outcome model or the propensity score model, and random-effects models additionally assume that the cluster effects are independent of the treatment and measured covariates. While subsequent propensity score methods \parencite{yang2018propensity, lee2021partially} improved these initial approaches, they still require parametric modeling assumptions, and thus remain susceptible to bias from model mis-specification. More recently, \Textcite{suk2022robust, suk2023tuning, suk2024within} developed robust ML methods to handle cluster-level unmeasured confounding. \Textcite{suk2022robust} introduced three ML-based estimators with ensemble supervised learning algorithms by minimizing new loss functions that are insensitive to unmeasured cluster-level  confounders. \Textcite{suk2023tuning} explored five modifications of existing ML methods, notably causal forests \parencite{wager2018estimation}, to enhance their robustness to cluster-level unmeasured confounding. \Textcite{suk2024within} proposed a within-group approach that groups similar clusters based on each cluster's treatment prevalence and trains ML methods within these groups to remove bias. 

Instead of modifying the estimation procedure, a few studies infer latent variables directly from observed data to robustly estimate causal effects. In particular, \textcite{morimura2025latent} introduced \emph{Latent DML}, which integrates latent variable modeling into the DML framework under the assumption of a constant treatment effect (via a partially linear regression model). The key idea is to compute outcome and treatment residuals using the standard DML nuisance estimation, fit a parametric latent variable model to these residuals via expectation-maximization (EM) to adjust the outcome residual, and then solve a score equation for causal effect estimation. In contrast, \textcite{louizos2017CEVAE} proposed a deep latent-variable model, named the Causal Effect Variational Autoencoder (CEVAE), to estimate treatment effects when proxy variables for unmeasured confounders are available. CEVAE uses a VAE architecture to simultaneously learn a latent representation of the unmeasured confounder and learn treatment and outcome models conditional on this latent variable; see Section \ref{sec:VAE} for details of VAE. Building on CEVAE, \textcite{Lu2020reconsidering} jointly learns a latent representation, the nuisance models, and the treatment effect model, with the R-learner loss in the variational objective. However, both \textcite{louizos2017CEVAE} and \textcite{Lu2020reconsidering} treat the observed covariates as noisy proxies of the latent confounder and are designed for independent and identically distributed (i.i.d.) data. Despite the increasing use of deep latent-variable models, there is limited research on inferring latent variables from residual structures in multilevel studies to develop robust ML methods.

The overarching goal of this paper is to propose a VAE-based approach for estimating the conditional average treatment effect (CATE) and the average treatment effect (ATE) in multilevel data with unmeasured cluster-level covariates in both observational studies and randomized trials. This approach leverages the within-cluster dependence of outcome and/or treatment residuals to infer latent confounders (or predictors) via amortized variational inference (i.e., efficient approximate posterior inference via an encoder network). It then adjusts the nuisance estimates accordingly and estimates the CATE using the R-learner loss function \parencite{nie2020quasi}, and the ATE using an augmented inverse propensity weighting (AIPW) estimator. Briefly, the R-learner estimates the CATE by regressing the residualized outcome on the residualized treatment, and serves as the underlying loss function for existing ML methods like causal forests; see Section \ref{sec:Rlearn} for details. We call our method the \emph{variational autoencoder-augmented R-learner} (VAE-R). \raggedbottom

Our contributions are fourfold. First, we repurpose amortized variational inference to infer unmeasured cluster-level confounders from within-cluster residual dependence, for robust CATE and ATE estimation in multilevel studies. Second, we specify the decoder through an offset likelihood, so that the latent variables capture only the cluster-level variation left unexplained by the measured covariates, and combine the resulting adjustments with cross-fitting and the R-learner objective. To the best of our knowledge, this is the first work to integrate an R-learner with VAEs in multilevel studies. Third, extending beyond confounders, we adapt our approach to handle unmeasured cluster-level predictors to achieve efficiency gains, an extension applicable to both randomized trials and observational studies. Fourth, we provide data-driven strategies for selecting the two hyperparameters of the VAE: the latent dimension size and the Kullback-Leibler (KL) weight. For the former, we adapt the automatic relevancy detection strategy of \textcite{saha2025ardvae} to our offset-likelihood setting. For the latter, we propose criteria based on the cluster-level dependence remaining in the adjusted residuals, tailored to observational studies and randomized trials, respectively.

\section{Setup}

\subsection{Notation and Causal Estimands}

Suppose we have $J$ clusters, indexed $j=1, \ldots, J$, where each cluster contains $n_j$ individuals, indexed by $i=1, \ldots, n_j$. The total sample size across all clusters is $n=\sum_{j=1}^{J} n_j$. For each individual $i$ in cluster $j$ (hereafter individual $ij$), we observe a binary treatment indicator $T_{ij} \in \{0, 1\}$, where $T_{ij}=1$ represents the treated status of individual $ij$ and $T_{ij}=0$ represents the untreated/control status for that individual. We denote the observed outcome for individual $ij$ as $Y_{ij}$. We define measured pre-treatment covariates at the individual and cluster levels as $X_{ij} \in \mathcal{X}$ and $W_{j} \in \mathcal{W}$, respectively. Unmeasured cluster-level covariates are denoted as $Z_j \in \mathcal{Z}$. We also use $C_{j} = j \in \{1, \ldots, J\}$ as an explicit cluster identifier. As a result, the complete set of measured and unmeasured variables for each individual is $O_{ij}=(Y_{ij}, T_{ij}, X_{ij}, W_j, Z_j, C_{j})$.

In this paper, we consider two main settings of interest: a multilevel observational study with unmeasured cluster-level confounders and a multisite randomized trial with unmeasured cluster-level predictors. While one can certainly imagine unmeasured cluster-level predictors within multilevel observational studies, our primary focus in the observational setting is the more severe threat of unmeasured cluster-level confounders. Figure \ref{fig:DAG} illustrates the causal diagrams for these two settings. In the multilevel observational study, measured individual-level covariates $X_{ij}$, measured cluster-level covariates $W_j$, and unmeasured cluster-level covariates $Z_j$ all serve as confounders affecting $T_{ij}$ and $Y_{ij}$. In contrast, in the multisite randomized trial, $X_{ij}, W_j$, and $Z_j$ serve as predictors of the outcome, and the unmeasured $Z_j$ does not confound the treatment. The measured cluster-level covariates $W_j$ may influence $T_{ij}$ if treatment assignment probabilities differ across clusters (e.g., urban schools assign students to treatment with a higher probability than rural schools). 

\begin{figure}[h]
\centering
    \subcaptionbox{Multilevel observational data \label{fig1}}[.42\textwidth]{
\begin{tikzpicture}
\tikzset{line width=1pt, outer sep=1pt,
ell/.style={draw,fill=white, inner sep=3pt,
line width=1pt},
ellempty/.style={inner sep=3pt,
line width=1pt}};
\node[name=A,ell,  shape=ellipse] at (-4,0) {$T_{ij}$}  ;
\node[name=Y,ell,  shape=ellipse] (Y) at (2,0) {$Y_{ij}$};
\node[name=X,ell,  shape=ellipse] at (-3,2) {$X_{ij}$};
\node[name=W,ell,  shape=ellipse] at (-1,2) {$W_j$};
\node[name=Z,ell,  shape=ellipse] at (1,2) {$Z_{j}$};

\begin{scope}[>={Stealth[black]},
      every edge/.style={draw=black,line width=0.5pt}]
\path [->] (A) edge (Y);
\path [->] (X) edge (Y);
\path [->] (X) edge (A);
\path [->] (W) edge (A);
\path [->] (W) edge (Y);
\path [->] (Z) edge (A);
\path [->] (Z) edge (Y);
\end{scope}
\end{tikzpicture}}
\hspace{0.1in}
    \subcaptionbox{Multisite randomized trial \label{fig2}}[.42\textwidth]{
\begin{tikzpicture}
\tikzset{line width=1pt, outer sep=1pt,
ell/.style={draw,fill=white, inner sep=3pt,
line width=1pt},
ellempty/.style={inner sep=3pt,
line width=1pt}};
\node[name=A,ell,  shape=ellipse] at (-4,0) {$T_{ij}$}  ;
\node[name=Y,ell,  shape=ellipse] (Y) at (2,0) {$Y_{ij}$};
\node[name=X,ell,  shape=ellipse] at (-3,2) {$X_{ij}$};
\node[name=W,ell,  shape=ellipse] at (-1,2) {$W_j$};
\node[name=Z,ell,  shape=ellipse] at (1,2) {$Z_{j}$};

\begin{scope}[>={Stealth[black]},
      every edge/.style={draw=black,line width=0.5pt}]
\path [->] (A) edge (Y);
\path [->] (X) edge (Y);
\path [->] (W) edge (A);
\path [->] (W) edge (Y);
\path [->] (Z) edge (Y);
\end{scope}
\end{tikzpicture}
}
\caption{Graphical representations of causal relationships in (a) a multilevel observational study and (b) a multisite randomized trial. $Z_j$ is unmeasured in both settings.}\label{fig:DAG}
\end{figure}

Additionally, we use the Neyman-Rubin potential outcome notation \parencite{neyman23, rubin1974} to define treatment effects. Let $Y_{ij}(1)$ represent the potential outcome that would occur if individual $ij$ were to receive the treatment (i.e., $T_{ij}=1$), and let $Y_{ij}(0)$ represent the potential outcome under the control condition ($T_{ij}=0$) for that individual. Using these potential outcomes, we define two causal estimands of interest: the ATE,
\[
\tau = \mathbb{E} \{ Y_{ij}(1) - Y_{ij}(0) \},
\]
and the CATE,
\[
\tau(x, w) = \mathbb{E} \{ Y_{ij}(1) - Y_{ij}(0) \mid X_{ij}=x, W_{j}=w \}.
\]
Here, the ATE represents the average difference between the two potential outcomes across the entire population, whereas the CATE represents the ATE among a specific subgroup of individuals defined by the measured pre-treatment covariates $X_{ij}$ and $W_{j}$.

\subsection{Causal Assumptions}

If the unmeasured covariates $Z_{j}$ were fully observed, the standard set of assumptions to identify the ATE and the CATE \parencite{imbens2015causal} in multilevel studies would be: 
\begin{itemize}
    \item[(A1)] \ \ \ \ Consistency: $Y_{ij}=Y_{ij}(t)$ if $T_{ij}=t$, for $t \in \{0, 1\}$
    \item[(A2)] \ \ \ \ Unconfoundedness: \\
       (i) for observational studies, $Y_{ij}(t) \indep T_{ij} \mid X_{ij}, W_j, Z_j \ \ \text{for} \ t \in \{0, 1\}$ \\
       (ii) for randomized trials, $Y_{ij}(t) \indep T_{ij} \mid W_j \ \ \text{for} \ t \in \{0, 1\}$
    \item[(A3)] \ \ \ \ Positivity: \\
       (i) for observational studies, $0 < e(X_{ij}, W_j, Z_j) = P(T_{ij} = 1 \mid X_{ij}, W_j, Z_j) < 1$ \\
       (ii) for randomized trials, $0 < e(W_j) = P(T_{ij} = 1 \mid W_j) < 1$   
\end{itemize}
Briefly, Assumption (A1) states that the observed outcome corresponds to the potential outcome under the actually assigned treatment status. It implies that an individual's potential outcomes are independent of others' treatment assignments (no interference) and that there are no multiple versions of the treatment. Assumption (A2)-(i) states that conditional on the full set of covariates $X_{ij}, W_j, Z_j$, the treatment assignment $T_{ij}$ behaves as if it were randomized, and is thus independent of $Y_{ij}(t)$. In contrast, (A2)-(ii) states that treatment is randomly assigned within each cluster, with assignment probabilities that may depend on $W_j$; this assumption is met by design in multisite randomized trials. Assumption (A3)-(i) states that  the probability of receiving treatment, i.e., the propensity score $e(X_{ij}, W_j, Z_j)$ is strictly between zero and one for every value of the covariates. In particular, (A3)-(i) rules out clusters where treatment assignment is deterministic so that every cluster can in principle contain both treated and control units. Assumption (A3)-(ii) states that the cluster-specific propensity score is strictly between zero and one, and is frequently known to researchers in experimental settings. 

While $Z_j$ does not bias treatment effect estimation in randomized trials, its role as an unmeasured confounder in observational studies violates the unconfoundedness assumption (A2)-(i) given the measured covariates alone. Therefore, standard identification strategies no longer work in observational data, and the ATE and the CATE are generally not identifiable without imposing additional structure on $Z_j$, or inferring it. 

\subsection{R-Learner}\label{sec:Rlearn}

There exist several meta-learners for CATE estimation, such as the Single (S)-learner, Two (T)-learner, Doubly Robust (DR)-learner, and R-learner \parencite{chernozhukov2024applied}. 
The R-learner relies on residualized estimation to control for observed confounding while allowing for flexible, nonparametric CATE estimation \parencite{nie2020quasi}. To outline the standard R-learner, we assume that there are no unmeasured covariates, so that (A2)-(i) holds given $X_{ij}$ and $W_j$. 

Let $m(x,w) = \mathbb{E}[Y_{ij} \mid X_{ij} = x, W_j=w]$ be the conditional mean of the outcome given specific values of covariates $x$ and $w$, and let $e(x,w)=P(T_{ij}=1 \mid X_{ij}=x, W_j=w)$ be the propensity score. Under (A1)--(A3), the Robinson decomposition yields $Y_{ij}-m(X_{ij}, W_j) = \tau(X_{ij}, W_j)\{T_{ij} - e(X_{ij}, W_j)\} + \epsilon_{ij}$, with $\mathbb{E}[\epsilon_{ij} \mid X_{ij}, W_j, T_{ij}] = 0$ \parencite{robinson1988root}, which motivates regressing the residualized outcome on the residualized treatment to estimate the CATE. 

The R-learner consists of a residualization step and an estimation step. First, we estimate the nuisance functions $m(x,w)$ and $e(x,w)$ using ML algorithms with cross-fitting. To accommodate the multilevel data structure, we partition the clusters into $K$ mutually exclusive folds. Let $\widehat{m}^{(-k)}(x, w)$ and $\hat{e}^{(-k)}(x, w)$ denote the nuisance functions estimates obtained using individuals from all clusters except those in fold $k$. For each observation in fold $k$, we compute the cross-fitted residuals: 
\[
\hat{U}_{ij}^{(k)}=Y_{ij} - \widehat{m}^{(-k)}(X_{ij}, W_j), \qquad  \hat{V}_{ij}^{(k)}=T_{ij} - \widehat{e}^{(-k)}(X_{ij}, W_j).
\]
Second, using the cross-fitted residuals across folds, we estimate the CATE by minimizing the R-learner objective:
\begin{align}
\hat{\tau} = \mathop{\text{argmin}}\limits_{\tau} \sum_{ij} \left( \hat{U}_{ij} - \tau(X_{ij}, W_j)\hat{V}_{ij} \right)^2
\end{align}
When $\hat{V}_{ij} \neq 0$, minimizing the R-learner loss is mathematically equivalent to minimizing a weighted squared loss, where the covariates are $X_{ij}, W_j$, the pseudo-outcome is $\hat{U}_{ij}/\hat{V}_{ij}$ and the observation weights are $\hat{V}_{ij}^2$ because $\{\hat{U}_{ij} - \tau(X_{ij}, W_j)\hat{V}_{ij}\}^2=$ $\hat{V}_{ij}^2 \{\hat{U}_{ij}/\hat{V}_{ij} - $ $\tau(X_{ij}, W_j)\}^2$. In practice, a regularization term can be added to the R-learner loss to control the functional complexity of $\tau(\cdot)$. 

The R-learner and DML may appear structurally similar in that both rely on cross-fitting and Neyman-orthogonal residualization to mitigate confounding. Despite these similarities, their goals differ. The R-learner is designed to model flexible, nonparametric forms of the CATE \parencite{nie2020quasi}, whereas traditional DML focuses on obtaining debiased, $\sqrt{n}$-consistent estimates and valid statistical inference for low-dimensional targets (e.g., the ATE or a parametric CATE) \parencite{chernozhukov2024applied}.

\subsection{Variational Autoencoders}\label{sec:VAE}

VAEs are deep latent-variable models that combine amortized variational inference with deep neural networks to learn low-dimensional probabilistic representations of complex data and generate new samples from these representations \parencite{kingma2014auto}. Conceptually, a VAE specifies a generative model $p_\theta(x, w, z) = p(z)p_\theta(x, w \mid z)$, where $p(z)$ is a prior distribution (commonly, a standard Gaussian) and $p_\theta(x, w \mid z)$ is a likelihood parameterized by $\theta$ using a decoder network. A VAE also introduces an encoder, or ``inference model'' $q_\phi(z\mid x, w)$, which approximates the generally intractable posterior $p_\theta(z \mid x, w)$. The encoder is a neural network parameterized by $\phi$ that specifies a tractable variational distribution, often a Gaussian with a diagonal covariance matrix.

VAEs are trained by maximizing the evidence lower bound (ELBO) on the log-marginal likelihood:
\begin{align}
\mathcal{L}(\theta,\phi; x, w) = \mathbb{E}_{q_{\phi}(z\mid x,w)}\!\left[\log p_{\theta}(x,w\mid z)\right]
-
\mathrm{KL}\!\left[q_{\phi}(z \mid x, w)\,\|\,p(z)\right] \leq \log p_{\theta}(x,w) 
\end{align}
Here, the first term is a reconstruction term that encourages accurate decoding of the observed data from the latent variables. The second term is a KL divergence that regularizes the approximate posterior toward the prior and yields a smooth, well-structured latent space. All parameters $\phi$ and $\theta$ are jointly optimized using stochastic gradient-based methods. Additionally, researchers often apply a weighting factor $\beta \geq 0$ to the KL term, yielding a $\beta$-weighted variational objective \parencite{higgins2017beta}. Compared to the standard ELBO ($\beta=1$), $\beta > 1$ strengthens the regularization toward the prior, whereas $0 < \beta < 1$ weakens it. To mitigate posterior collapse, researchers may employ KL annealing, gradually increasing the weight from zero to a target value ($\beta_{max}$) during training \parencite{bowman2016generating}.

Furthermore, when $q_{\phi}(z\mid x,w)$ is chosen as a Gaussian with mean $\mu_\phi(x,w)$ and diagonal standard deviation $\sigma_\phi(x,w)$, VAEs utilize the reparameterization trick: $z=\mu_\phi(x,w) +  \sigma_\phi(x,w) \odot \epsilon$, $\ \epsilon \sim \mathcal{N}(0,I)$. This is necessary because gradients cannot be directly backpropagated through the stochastic sampling of $z$. The reparameterization ensures that
\[
\mathbb{E}_{q_{\phi}(z\mid x,w)}\!\left[\log p_{\theta}(x,w\mid z)\right]
=
\mathbb{E}_{\epsilon \sim \mathcal{N}(0,I)}\!\left[
\log p_{\theta}\!\left(x,w \mid \mu_{\phi}(x,w) + \sigma_{\phi}(x,w)\odot \epsilon\right)
\right].
\]
which can be efficiently approximated using Monte Carlo samples $\epsilon^{(s)} \sim \mathcal{N}(0,I)$. Once trained, VAEs support both amortized inference and sampling-based generation by drawing $z$ from the prior and decoding it into synthetic data \parencite{diederik2019introduction}. In this paper, we integrate the architectural strengths of both the VAE and the R-learner to develop our robust causal method, VAE-R, for multilevel studies.

\section{VAE-R for Multilevel Studies with Unmeasured Covariates}

In this section, we propose VAE-R for estimating the CATE and ATE in multilevel data when unmeasured cluster-level covariates are present. These unmeasured covariates may serve as confounders in observational settings or as outcome predictors in randomized trials. At a high level, VAE-R utilizes within-cluster dependence in the residualized outcome and/or treatment to infer cluster-level latent variables via amortized variational inference. The decoder is specified through an offset likelihood so that the latent variables explain only the variation left unexplained by the initial nuisance models. Using the learned latent adjustments, we update the nuisance estimates for the outcome and, when needed, the treatment model, and then estimate the CATE by minimizing an R-learner objective. We then estimate the ATE by aggregating the estimated CATE via an influence-function-based (AIPW) estimator. See Figure \ref{fig:overview} for an illustration of VAE-R.

\begin{figure}[h]
\centering
\begin{tikzpicture}[
  node distance=1.1cm,
  box/.style={draw, rounded corners=2pt, minimum width=6.4cm, minimum height=0.95cm,
              align=center, font=\small, line width=0.8pt},
  vaebox/.style={box, fill=blue!6},
  flow/.style={->, >=Stealth, line width=1pt},
  note/.style={font=\scriptsize\itshape, align=left}
]
\node[box] (n0) {Nuisance Estimation\\[-1pt]{\scriptsize $\widehat m^{(0)}(X,W),\ \widehat e^{(0)}(X,W)$}};
\node[box, below=of n0] (n1) {Naive Residuals\\[-1pt]{\scriptsize $U_{ij}=Y_{ij}-\widehat m^{(0)}_{ij},\ \ V_{ij}=T_{ij}-\widehat e^{(0)}_{ij}$}};
\node[vaebox, below=of n1] (n2) {VAE: Latent Inference\\[-1pt]{\scriptsize $q_\phi(Z_j\mid S_j)$}};
\node[vaebox, below=of n2] (n3) {Residual Adjustment\\[-1pt]{\scriptsize $\widehat m^{*}_{ij}=\widehat m^{(0)}_{ij}+\widehat\delta_Y(\widehat Z_j)$}\\[-3pt]{\scriptsize $\widehat e^{*}_{ij}=\sigma\{\mathrm{logit}(\widehat e^{(0)}_{ij})+\widehat\delta_A(\widehat Z_j)\}$}};
\node[box, below=of n3] (n4) {R-learner: CATE Estimation\\[-1pt]{\scriptsize $\widehat\tau(X_{ij},W_j)$}};

\draw[flow] (n0) -- (n1);
\draw[flow] (n1) -- node[right, note] {\ cluster summaries $S_j$} (n2);
\draw[flow] (n2) -- node[right, note] {\ $\widehat Z_j,\ \widehat \delta_Y,\ \widehat \delta_A$} (n3);
\draw[flow] (n3) -- node[right, note] {\ $\widehat U^{*}_{ij},\ \widehat V^{*}_{ij}$} (n4);

\draw[flow, red, dashed]
  (n1.west) -- ++(-1.5,0)
  node[midway, above, font=\scriptsize\color{red}] {}
  |- (n4.west)
  node[pos=0.25, left, font=\scriptsize\color{red}, align=center] {Standard\\R-learner};

\end{tikzpicture}
\caption{Overview of VAE-R. The standard R-learner passes the naive residuals directly to the CATE estimation step (dashed path), whereas VAE-R inserts a latent inference and residual adjustment stage (shaded) that corrects the nuisance estimates for unmeasured cluster-level covariates. In the multisite randomized trial setting, only the outcome decoder $\delta_Y$ is used and $\widehat V_{ij}$
is left unadjusted.}\label{fig:overview}
\end{figure}

Below, we first introduce the version of VAE-R for handling unmeasured cluster-level confounders in multilevel observational data (Section \ref{sec:VAE_con}), and then describe a simplified version tailored to unmeasured cluster-level predictors in multisite randomized trials (Section \ref{sec:VAE_pred}). Both versions involve two hyperparameters, the latent dimension size $d_z$ and the KL weight $\beta$; Section \ref{sec:hyper} describes our data-driven strategies for selecting them.

\subsection{Unmeasured Cluster-Level Confounders}\label{sec:VAE_con}

The proposed VAE-R approach is motivated by the latent DML method \parencite{morimura2025latent} and the R-learner \parencite{nie2020quasi}. Consider the following structural equations for the outcome and the treatment:
\begin{align*}
    Y_{ij} &= \mu(X_{ij}, W_j) + \delta_Y(Z_j) + \tau(X_{ij}, W_j)T_{ij} + \epsilon_{ij}, \qquad \mathbb{E}[\epsilon_{ij} \mid X_{ij}, W_j, Z_j, T_{ij}] = 0 \\
    T_{ij} &\sim \text{Bernoulli}\big(\sigma( \text{logit}(e(X_{ij}, W_j)) + \delta_A(Z_j)) \big),
\end{align*}
where $\tau(X, W)$ is the CATE of interest, $\mu(X, W)$ is the outcome baseline under control, and $e(X, W)$ is the baseline propensity score based on measured covariates. The functions $\delta_Y(Z_j)$ and $\delta_A(Z_j)$ represent the additive shifts introduced by the unmeasured confounder $Z_j$ in the outcome model and the treatment model (on the logit scale), respectively, and $\sigma(\cdot)$ denotes the sigmoid function. 

Similarly to standard and latent DML approaches, we first estimate the initial, naive nuisance functions using only the measured covariates, $m^{(0)}(x, w)=\mathbb{E}[Y_{ij} \mid X_{ij}=x, W_j=w]$ and $e^{(0)}(x, w)= \mathbb{E}[T_{ij} \mid X_{ij}=x, W_j=w]$. In our implementation, we obtain $\widehat{m}^{(0)}$ by fitting separate outcome regressions within each treatment arm, $\widehat{g}_t(x,w)$ for $t \in \{0, 1\}$, and combining them using the estimated propensity score, $\widehat{m}^{(0)} = \hat{e}^{(0)}\widehat{g}_1 + (1-\hat{e}^{(0)})\widehat{g}_0$. We then compute the initial residualized outcome and residualized treatment as follows: 
\[
\widehat{U}_{ij}=Y_{ij}-\widehat{m}^{(0)}(X_{ij}, W_j), \qquad \widehat{V}_{ij}=T_{ij}-\widehat{e}^{(0)}(X_{ij}, W_j).  
\]
Because $Z_{j}$ is omitted from the nuisance models, individuals' residuals in the same cluster are dependent on one another. This within-cluster dependence is the signal that the encoder extracts from the cluster summaries below. 

Using these initial residuals, we then infer a low-dimensional representation of $Z_j$ via a VAE. Specifically, we use cluster-level summaries $S_j \in \mathbb{R}^s$, instead of the raw individual-level residuals, because a fixed-dimensional encoder input is stable and computationally efficient. We construct $S_j$ using the cluster means and variances of each residual, the cluster mean of their products, and the cluster mean of absolute residuals like: 
\[
S_j = \left(\overline{\widehat{U}}_j, \overline{\widehat{V}}_j, \text{Var}(\widehat{U})_j, \text{Var}(\widehat{V})_j,  \overline{\widehat{V}\widehat{U}}_j, \overline{|\widehat{V}|}_j \right).
\]
For numerical stability, we compute these summaries after standardizing ($\widehat{U}, \widehat{V}$), and pass them as inputs to the encoder.

The encoder $q_\phi$ defines an approximate posterior for each cluster as:
\[
q_\phi(Z_j \mid S_j) = \mathcal{N}\left(\mu_\phi(S_j), \text{diag}(\sigma^2_\phi(S_j)) \right).
\]
where $\mu_\phi(\cdot)$ and $\log\sigma^2_\phi(\cdot)$ are the outputs of a feed-forward neural network with two hidden layers. The decoder $p_\theta$ is constrained to model the data as deviations from the naive offsets. Specifically, $\delta_Y$ and $\delta_A$ are cluster-level additive corrections; in our implementation, we use linear decoders, $\delta_Y(Z_j) = b_Y^\top Z_j$ and $\delta_A(Z_j) = b_A^\top Z_j$, with loading vectors $b_Y, b_A \in \mathbb{R}^{d_z}$. The corresponding offset likelihoods are:
\begin{align}
Y_{ij} &\mid Z_j \sim \mathcal{N}(\widehat{m}^{(0)}_{ij} + \delta_Y(Z_j), \sigma^2_Y), \\
T_{ij} &\mid Z_j \sim \text{Bernoulli}\left(\sigma( \text{logit}(\hat{e}^{(0)}_{ij}) + \delta_A(Z_j) ) \right),
\end{align}
where the initial nuisance predictions enter the decoder as known offsets and $\sigma^2_Y$ is estimated jointly with $b_Y$ and $b_A$. This architecture encourages the latent variable $Z_j$ to capture only the residual variation left unexplained by $X$ and $W$, preventing the VAE from relearning known covariate effects. As a result, we train VAE by minimizing the negative ELBO as:
\begin{align}
\mathcal{L}(\phi, \theta; S, \beta) = -\sum_{j=1}^J \mathbb{E}_{q_\phi (Z_j | S_j)} \left[ \sum_{i=1}^{n_j}  \big\{ \log p(Y_{ij} \mid Z_j) +  \log p(T_{ij} \mid Z_j)  \big\} \right] + \beta \sum_{j=1}^J \text{KL}\left(q_\phi(Z_j \mid S_j) \, \|\, \mathcal{N}(0, I) \right). \label{eq:ELBO}
\end{align}
Here, $\beta > 0$ controls the strength of the KL regularization and governs the trade-off between  reconstruction accuracy (data likelihood) and adherence to the prior distribution. Increasing $\beta$ enforces stronger shrinkage of $q_\phi(Z_j \mid S_j)$ toward $\mathcal{N}(0, I)$, whereas decreasing $\beta$ allows the latent representation to fit the data more closely. Since the reconstruction term sums $n_j$ individual contributions but the KL term contributes once per cluster, the effective strength of the regularization depends on the cluster size, as well as the strength of the cluster-level signal, and the latent dimension size. Thus, no single value of $\beta$ is appropriate across datasets, and we treat $\beta$ as a tunable hyperparameter (see Section \ref{sec:hyper}). To mitigate posterior collapse and stabilize optimization, we use KL warmup: we replace $\beta$ with $\beta^{(r)}=\min(\beta, \beta\frac{r}{\rho R})$, $r=0,1,\ldots, R-1$, which increases linearly from 0 to $\beta$ over the first fraction $\rho$ of the $R$ training epochs. We optimize \eqref{eq:ELBO} using Adam \parencite{kingma2015adam} with full-batch updates (i.e., one gradient step per epoch).

After fitting the VAE, we obtain posterior mean estimates $\widehat{Z}_j=\mu_{\widehat{\phi}}(S_j)$ and decoder estimates $\widehat{\delta}_Y$ and $\widehat{\delta}_A$, and compute the adjusted residuals:
\begin{align}
\widehat{U}^*_{ij} &= Y_{ij} - \widehat{m}^{*}_{ij}, \qquad \widehat{m}^{*}_{ij} = \widehat{m}^{(0)}_{ij} + \widehat{\delta}_Y (\widehat{Z}_j), \\
\widehat{V}^*_{ij} &= T_{ij} -  \widehat{e}^*_{ij}, \qquad 
\widehat{e}^*_{ij} = \sigma\big( \text{logit}(\widehat{e}^{(0)}_{ij}) + \widehat{\delta}_A (\widehat{Z}_j) \big).
\end{align}
We then estimate the CATE $\tau(X, W)$ by minimizing the R-learner objective using these adjusted residuals:
\[
\hat{\tau} = \mathop{\text{argmin}}\limits_{\tau} \frac{1}{n} \sum_{ij}\{\widehat{U}^*_{ij} - \tau(X_{ij}, W_j)\widehat{V}^*_{ij}\}^2.
\]
This objective can be minimized using any regression learner for $\tau(\cdot)$. Specifically, we use a weighted regression of $\widehat{U}^*_{ij}/\widehat{V}^*_{ij}$ on ($X_{ij}, W_j$) with sample weights $\widehat{V}^{*2}_{ij}$, which is mathematically equivalent to the above objective. Once the CATE is estimated, we estimate the ATE by aggregating the estimated CATEs. In particular, we use the sample average operator adapted to the cluster structure: 
\begin{align}\label{eq:naiveATE}
\hat{\tau}_{\rm ATE} = \frac{1}{J} \sum_{j=1}^{J} \frac{1}{n_j} \sum_{i=1}^{n_j} \hat{\tau}(X_{ij}, W_j) 
\end{align}
We also use the AIPW estimator motivated by the efficient influence-function representation: 
\begin{align}\label{eq:AIPW}
\hat{\tau}_{\rm ATE, AIPW} =  \frac{1}{J} \sum_{j=1}^{J} \frac{1}{n_j} \sum_{i=1}^{n_j} \psi_{ij}, \quad \psi_{ij} = \hat{\tau}(X_{ij}, W_j) + \frac{T_{ij} - \hat{e}^*_{ij}}{\hat{e}^*_{ij}(1-\hat{e}^*_{ij})}\big( (Y_{ij} - \widehat{m}^*_{ij}) - \hat{\tau}(X_{ij}, W_j)(T_{ij}-\hat{e}^*_{ij})\big).
\end{align}
For statistical inference on the ATE and CATE, we employ a cluster bootstrap that resamples clusters, rather than individuals, to estimate their standard errors and confidence intervals.

\begin{algorithm}[!b]
\caption{\small{VAE-R for Unmeasured Cluster-Level Confounders in Multilevel Observational Data}}\label{algo1}
\begin{algorithmic}[1]
\small
\Require Outcome $Y_{ij}$, treatment $T_{ij}$, covariates $X_{ij}, W_j$, cluster IDs $j \in \{1,\dots,J\}$, number of folds $K$, inner folds $K_{\rm inner}$, KL weight $\beta$, latent dimension $d_z$.
\State Partition clusters into $K$ synchronized folds $\mathcal{C}_1, \dots, \mathcal{C}_K$.
\State Let $\mathcal{I}^{(-k)}=\{(i,j): j\notin\mathcal{C}_k\}$ (training set) and $\mathcal{I}^{(k)}=\{(i,j): j\in\mathcal{C}_k\}$ (test set).
\For{$k = 1$ to $K$}
    \State \textbf{Stage 1 (Naive Nuisances):} Fit $\widehat{m}^{(0)}$ and $\widehat{e}^{(0)}$ on $\mathcal{I}^{(-k)}$  and predict $\widehat{m}^{(0)}_{ij}$ and $\widehat{e}^{(0)}_{ij}$ for $ij \in \mathcal{I}^{(k)}$; obtain $\hat{m}^{(0)}_{ij}$  and $\widehat{e}^{(0)}_{ij}$ for $ij \in \mathcal{I}^{(-k)}$ by $K_{\rm inner}$-fold cluster cross-fitting within $\mathcal{I}^{(-k)}$.  
    \State Compute naive residuals $\widehat{U}_{ij} = Y_{ij} - \hat{m}^{(0)}_{ij}$ and $\widehat{V}_{ij} = T_{ij} - \hat{e}^{(0)}_{ij}$ for all $ij$, and standardize them using the mean and standard deviation on $\mathcal{I}^{(-k)}$.
    \State \textbf{Stage 2 (Latent Inference):} Compute cluster summaries $S_j$ from the standardized residuals for all $j$. Train the VAE on $\mathcal{I}^{(-k)}$ to learn the encoder $q_\phi(Z|S)$ and the decoder $p_\theta$ with offsets and $\delta_Y, \delta_A$ by minimizing the negative ELBO:
  \[
    -\sum_{j\notin\mathcal{C}_k}\mathbb{E}_{q_\phi(Z_j\mid S_j)}\!\left[\sum_{i=1}^{n_j}\{\log p(Y_{ij}\mid Z_j) + \log p(T_{ij}\mid Z_j)\}\right]
    +\beta \sum_{j\notin\mathcal{C}_k}\mathrm{KL}\!\left(q_\phi(Z_j\mid S_j)\,\|\,\mathcal{N}(0,I)\right).
  \]
    
    \State Infer posterior means $\hat{Z}_j$ for all clusters in both $\mathcal{I}^{(-k)}$ and $\mathcal{I}^{(k)}$.
    \State \textbf{Stage 3 (Residual Adjustment):} Compute adjusted residuals for all $ij$:
        \begin{align}
        \widehat{U}^*_{ij} &= Y_{ij} - \widehat{m}^{*}_{ij}, \qquad \widehat{m}^{*}_{ij} = \widehat{m}^{(0)}_{ij} + \widehat{\delta}_Y (\widehat{Z}_j), \\
        \widehat{V}^*_{ij} &= T_{ij} -  \widehat{e}^*_{ij}, \qquad 
        \widehat{e}^*_{ij} = \sigma\big( \text{logit}(\widehat{e}^{(0)}_{ij}) + \widehat{\delta}_A (\widehat{Z}_j) \big).
        \end{align}
    \State \textbf{Stage 4 (CATE Estimation):} Estimate $\hat{\tau}^{(-k)}(X_{ij}, W_j)$ by minimizing the R-learner objective on $\mathcal{I}^{(-k)}$:
    $$ \hat{\tau}^{(-k)} = \mathop{\text{argmin}}\limits_{\tau} \sum_{ij \in \mathcal{I}^{(-k)}} \left( \widehat{U}^*_{ij} - \tau(X_{ij}, W_j) \widehat{V}^*_{ij}  \right)^2 $$
    \State Predict the CATE $\hat{\tau}(X_{ij}, W_j)=\hat{\tau}^{(-k)}(X_{ij}, W_j)$ and store $\widehat{m}^{*}_{ij}$, $\widehat{e}^{*}_{ij}$ for $ij \in \mathcal{I}^{(k)}$.
\EndFor
\State Compute the ATE estimates from the out-of-fold quantities: the sample average $\hat{\tau}_{\rm ATE}$ in \eqref{eq:naiveATE} and the AIPW estimator $\hat{\tau}_{\rm ATE, AIPW}$ in \eqref{eq:AIPW}.

\Ensure Out-of-fold CATE estimates $\hat{\tau}(X_{ij}, W_j)$ for all $ij$, out-of-fold nuisances $\widehat{m}^{*}_{ij}$ and $\widehat{e}^{*}_{ij}$, and ATE estimates $\hat{\tau}_{\rm ATE}$ and $\hat{\tau}_{\rm ATE, AIPW}$.
\end{algorithmic}
\end{algorithm} \raggedbottom

Algorithm \ref{algo1} summarizes the procedure of our approach for a given KL weight $\beta$ and latent dimension $d_z$. For implementation, we utilize cluster-level cross-fitting  to avoid cluster leakage and reduce overfitting bias. Specifically, we partition the clusters into $K$ folds, $\mathcal{C}_1, \dots, \mathcal{C}_K$. For each fold $k$, all nuisance and latent components are trained on the out-of-fold clusters, denoted as $\mathcal{I}^{(-k)}=\{(i,j): j\notin\mathcal{C}_k\}$, and treatment effect predictions are evaluated on the held-out clusters, denoted as $\mathcal{I}^{(k)}=\{(i,j): j\in\mathcal{C}_k\}$. Within each fold, the naive nuisance models fitted on $\mathcal{I}^{(-k)}$ supply the offsets $\widehat{m}^{(0)}_{ij}$ and $\widehat{e}^{(0)}_{ij}$ for the held-out clusters, whereas the training clusters' offsets are obtained by a nested $K_{\rm inner}$-fold cluster split of $\mathcal{I}^{(-k)}$, so that no training individual's own outcome or treatment contributes to its offset. The VAE and the CATE learner are then trained on  $\mathcal{I}^{(-k)}$, and the fitted encoder is applied to the held-out clusters in $\mathcal{I}^{(k)}$ to obtain their latent adjustments and CATE predictions.

\subsection{Unmeasured Cluster-Level Predictors}\label{sec:VAE_pred}

We now turn to the presence of unmeasured cluster-level predictors in multisite randomized trials. Consider the following structural equations for the outcome and the treatment, where $Z_j$ serves as an unmeasured cluster-level predictor: 
\begin{align*}
    Y_{ij} &= \mu(X_{ij}, W_j) + \delta_Y(Z_j) + \tau(X_{ij}, W_j)T_{ij} + \epsilon_{ij}, \qquad \mathbb{E}[\epsilon_{ij} \mid X_{ij}, W_j, Z_j, T_{ij}] = 0 \\
    T_{ij} &\sim \text{Bernoulli}\big(e(W_j)\big),
\end{align*}
where $\tau(X, W)$ is the CATE of interest, $\mu(X, W)$ is the outcome baseline under control, and $e(W)$ is the known propensity score, which may depend on cluster-level covariates. If it does, clusters have different treatment assignment probabilities. The function $\delta_Y(Z_j)$ represents the additive shift induced by the unmeasured predictor $Z_j$ in the outcome model. 

We first estimate the naive outcome nuisance function using only the measured covariates, $m^{(0)}(x, w)=\mathbb{E}[Y_{ij} \mid X_{ij}=x, W_j=w]$, again as the propensity-weighted combination of arm-specific regressions. For the treatment model, we use the known propensity score $e(W_j)$, which can also be estimated as the observed proportion of treated units within each cluster. We then compute the initial residualized outcome and residualized treatment as follows: 
\[
\widehat{U}_{ij}=Y_{ij}-\widehat{m}^{(0)}(X_{ij}, W_j), \qquad \widehat{V}_{ij}=T_{ij}-\widehat{e}(W_j).  
\]
Because the unmeasured variable $Z_j$ serves as a predictor of the outcome and does not confound treatment assignment, the treatment residual $\widehat{V}_{ij}$ contains no information about $Z_j$. Therefore, while $\widehat{V}_{ij}$ is required for the final R-learner stage, only the outcome residual $\widehat{U}_{ij}$ is used to infer the latent structure in the downstream VAE stage. 

More specifically, the cluster-level summaries $S_j$ used in the encoder are computed from the (standardized) outcome residual only. For example, we use the cluster mean, variance, mean absolute value, mean square, and mean sign of $\widehat{U}$: 
\[
S_j = \left(\overline{\widehat{U}}_j, \text{Var}(\widehat{U})_j, \overline{|\widehat{U}|}_j, \overline{\widehat{U}^2}_j, \overline{\text{sign}(\widehat{U})}_j \right).
\]
The decoder then reconstructs only the outcome with the naive outcome offset: 
\begin{align}
Y_{ij} &\mid Z_j \sim \mathcal{N}(\widehat{m}^{(0)}_{ij} + \delta_Y(Z_j), \sigma^2_Y). \end{align}
and the negative ELBO for VAE training simplifies to:
\[
\mathcal{L}(\phi, \theta; S, \beta) = -\sum_{j=1}^J \mathbb{E}_{q_\phi (Z_j | S_j)} \left[ \sum_{i=1}^{n_j}  \log p(Y_{ij} \mid Z_j) \right] + \beta \sum_{j=1}^J \text{KL}\left(q_\phi(Z_j \mid S_j) \, \|\, \mathcal{N}(0, I) \right).
\]
In this setting, we additionally employ a leave-one-out (LOO) inference scheme for the latent predictor. While the encoder is trained using the full-cluster summaries $S_j$ as above, for each individual $ij$ we infer the latent predictor from the summary that excludes that individual's own residual, 
\[
\widehat{Z}_{j,-i} = \mu_{\widehat{\phi}}(S_{j,-i}), \quad S_{j,-i} = \left(\overline{U}_{j,-i}, \text{Var}(U)_{j,-i}, \overline{|U|}_{j,-i}, \overline{U^2}_{j,-i}, \overline{\text{sign}(U)}_{j,-i} \right) \ ,
\]
where each summary is computed over the other $n_j-1$ individuals in cluster $j$. This ensures that the encoder input used to construct the adjusted outcome nuisance $\widehat{m}^*_{ij}$ does not contain $Y_{ij}$ itself, so that $\widehat{m}^*_{ij}$ is not fitted to the individual's own outcome.\footnote{In the confounder setting, we use the full-cluster summaries to infer the shared latent factor, because LOO inference weakens the joint treatment-outcome signal required for confounder recovery and can leave residual confounding.} The adjusted outcome residual and the treatment residual are as follows:
\begin{align}
\widehat{U}^*_{ij} &= Y_{ij} - \widehat{m}^*_{ij}, \qquad \widehat{m}^*_{ij}=\left( \widehat{m}^{(0)}_{ij} + \widehat{\delta}_Y (\widehat{Z}_{j,-i}) \right), \\
\widehat{V}_{ij} &= T_{ij} -  \widehat{e}(W_j).
\end{align}
We then estimate the CATE $\tau(X, W)$ by minimizing the R-learner objective using these residuals:
\[
\hat{\tau} = \mathop{\text{argmin}}\limits_{\tau} \frac{1}{n} \sum_{ij}\{\widehat{U}^*_{ij} - \tau(X_{ij}, W_j)\widehat{V}_{ij}\}^2.
\]
Additionally, we estimate the ATE using the sample average and the AIPW estimator as above, with $\widehat{e}(W_j)$ in place of $\widehat{e}^*_{ij}$, and obtain standard errors and confidence intervals by the cluster bootstrap. The specific procedure for the presence of unmeasured cluster-level predictors is summarized in Algorithm \ref{algo2}, with cluster-level cross-fitting.

\begin{algorithm}[!tb]
\caption{\small{VAE-R for Unmeasured Cluster-Level Predictors in Multisite Randomized Trials}}\label{algo2}
\begin{algorithmic}[1]
\small
\Require Outcome $Y_{ij}$, treatment $T_{ij}$, covariates $X_{ij}, W_j$, propensity score $e(W_j)$ (known or estimated as the within-cluster treated proportion), cluster IDs $j \in \{1,\dots,J\}$, number of folds $K$, inner folds $K_{\rm inner}$, KL weight $\beta$, latent dimension $d_z$.
\State Partition clusters into $K$ synchronized folds $\mathcal{C}_1, \dots, \mathcal{C}_K$.
\State Let $\mathcal{I}^{(-k)}=\{(i,j): j\notin\mathcal{C}_k\}$ (training set) and $\mathcal{I}^{(k)}=\{(i,j): j\in\mathcal{C}_k\}$ (test set)
\For{$k = 1$ to $K$}
    \State \textbf{Stage 1 (Naive Nuisances):} Fit $\hat{m}^{(0)}$ on $\mathcal{I}^{(-k)}$ using ($X, W$) and predict $\hat{m}^{(0)}_{ij}$ for $ij \in \mathcal{I}^{(k)}$; obtain $\hat{m}^{(0)}_{ij}$ for $ij \in \mathcal{I}^{(-k)}$ by $K_{\rm inner}$-fold cluster cross-fitting within $\mathcal{I}^{(-k)}$. 
    \State Compute naive residuals $\widehat{U}_{ij} = Y_{ij} - \hat{m}^{(0)}_{ij}$ and $\widehat{V}_{ij} = T_{ij} - \hat{e}(W_j)$ for all $ij$, and standardize $\widehat{U}$ using the mean and the standard deviation on $\mathcal{I}^{(-k)}$.
    \State \textbf{Stage 2 (Latent Inference):} Compute cluster summaries $S_j$ from the standardized outcome residuals for all $j$.
    \State Train the VAE on $\mathcal{I}^{(-k)}$ to learn the encoder $q_\phi(Z|S)$ and the decoder $p_\theta$ with the offset and $\delta_Y$ by minimizing the negative ELBO:
  \[
    -\sum_{j\notin\mathcal{C}_k}\mathbb{E}_{q_\phi(Z_j\mid S_j)}\!\left[\sum_{i=1}^{n_j}\log p(Y_{ij}\mid Z_j)\right]
    +\beta \sum_{j\notin\mathcal{C}_k}\mathrm{KL}\!\left(q_\phi(Z_j\mid S_j)\,\|\,\mathcal{N}(0,I)\right).
  \]
    
    \State Compute LOO summaries $S_{j,-i}$ and infer $\widehat{Z}_{j,-i} = \mu_{\hat{\phi}}(S_{j,-i})$ for all $ij$ in both $\mathcal{I}^{(-k)}$ and $\mathcal{I}^{(k)}$.
    \State \textbf{Stage 3 (Residual Adjustment):} Compute the adjusted outcome residual for all $ij$:
        \begin{align}
        \widehat{U}^*_{ij} &= Y_{ij} - \widehat{m}^{*}_{ij}, \qquad \widehat{m}^{*}_{ij} = \widehat{m}^{(0)}_{ij} + \widehat{\delta}_Y (\widehat{Z}_{j,-i}) 
        \end{align}
    \State \textbf{Stage 4 (CATE Estimation):} Estimate $\hat{\tau}^{(-k)}(X_{ij}, W_j)$ by minimizing the R-learner objective on $\mathcal{I}^{(-k)}$:
    $$ \hat{\tau}^{(-k)} = \mathop{\text{argmin}}\limits_{\tau} \sum_{ij \in \mathcal{I}^{(-k)}} \left( \widehat{U}^*_{ij} - \tau(X_{ij}, W_j) \widehat{V}_{ij}  \right)^2 $$
    \State Predict the CATE $\hat{\tau}(X_{ij}, W_j)=\hat{\tau}^{(-k)}(X_{ij}, W_j)$ and store $\widehat{m}^*_{ij}$ for $ij \in  \mathcal{I}^{(k)}$.
\EndFor
\State Compute the ATE estimates from the out-of-fold quantities: the sample average $\hat{\tau}_{\rm ATE}$ in \eqref{eq:naiveATE} and the AIPW estimator $\hat{\tau}_{\rm ATE, AIPW}$ in \eqref{eq:AIPW}, with $\widehat{e}(W_j)$ in place of $\widehat{e}^*_{ij}$.
\Ensure Out-of-fold CATE estimates $\hat{\tau}(X_{ij}, W_j)$ for all $ij$, out-of-fold nuisance $\widehat{m}^*_{ij}$ and propensity $\widehat{e}(W_j)$, and ATE estimates $\hat{\tau}_{\rm ATE}$ and $\hat{\tau}_{\rm ATE, AIPW}$.
\end{algorithmic}
\end{algorithm} 

\subsection{Selection of the KL Weight and the Latent Dimension Size}\label{sec:hyper}

Algorithms \ref{algo1} and \ref{algo2} take the KL weight $\beta$ and the latent dimension size $d_z$ as given. In practice, the true dimension of $Z_j$ is unknown, and the appropriate amount of KL regularization depends on the cluster sizes. We select the two hyperparameters jointly through a nested procedure. For each candidate $\beta$ in a user-specified grid $\mathcal{B}$, we first determine a data-driven latent dimension $d_z(\beta)$ using an automatic relevancy detection (ARD) strategy, refit VAE-R at ($\beta, d_z(\beta)$), and then select $\beta$ using an observable diagnostic based on the out-of-fold residuals. We describe the two components in turn, and Algorithm \ref{algo3} summarizes the full procedure.

\subsubsection{Latent dimension size via ARD} To determine $d_z$ for a given $\beta$, we adapt the ARD-VAE of \textcite{saha2025ardvae}. The key idea is simple. In the ordinary VAE, every latent axis is assigned the same prior variance of one, so the model has no way to signal that some axes are unnecessary. ARD instead lets each axis have its own prior variance that is learned from the data during training. That is, an axis that the encoder actually uses to distinguish clusters obtains a sizable variance, whereas an axis that carries no information is shrunk toward zero. After training, the axes whose variance has collapsed can be discarded, and the number of surviving axes serves as the selected latent dimension. 

Specifically, we start from a generous ceiling $d_{\rm max}$ on the number of latent axes and replace the fixed prior $\mathcal{N}(0, I)$ with $\mathcal{N}(0, \text{diag}(\hat{\alpha}^{-1}))$, where $\hat{\alpha}^{-1}_l$ is the prior variance of axis $l$.\footnote{Following \textcite{saha2025ardvae}, the prior $\mathcal{N}(0, \text{diag}(\hat{\alpha}^{-1}))$ corresponds to placing independent Gamma priors on the precisions, $\alpha_l \sim \mathrm{Gamma}(a_l^0,b_l^0)$, with noninformative prior $a_l^0=b_l^0=0$, and approximating the resulting Student-$t$ marginal prior on $Z_j$ by a Gaussian.} The variances are re-estimated each epoch in a straightforward way; we draw one latent value $z_{jl}$ from the encoder for each cluster in a randomly selected subset $\mathcal{S}_\alpha$ (a fraction $\rho_\alpha$ of the training clusters), and set $\hat{\alpha}^{-1}_l$ to the mean of $z^2_{jl}$ across those clusters. The encoder and decoder are then updated by one gradient step on the negative ELBO computed on the remaining clusters $\mathcal{S}_{\rm sgd}$, using the same KL weight $\beta$ and warmup schedule as in Section \ref{sec:VAE_con}; the only change to the ELBO is that the KL term measures the distance from $q_\phi(Z_j \mid S_j)$ to the Gaussian prior with axis-specific variances rather than to $\mathcal{N}(0, I)$. Holding out $\mathcal{S}_\alpha$ for the variance update and using $\mathcal{S}_{\rm sgd}$ for the gradient step prevents the same clusters from setting the prior and being regularized toward it within an epoch.

Once training is complete, we score each axis by how much it matters to the reconstruction. A large prior variance alone is not sufficient, because an axis can vary across clusters yet have negligible effect on the decoded outcome and treatment. Therefore, we weight each variance by the squared decoder loadings on that axis,
\[
\text{rel}_l = \big(\widehat{b}_{Y,l}^2 + \widehat{b}_{A,l}^2\big)\,\widehat{\alpha}_l^{-1},
\qquad l = 1, \ldots, d_{\max},
\]
where $\widehat{b}_{Y,l}$ and $\widehat{b}_{A,l}$ are the $l$-th elements of the estimated loading vectors $\widehat{b}_{Y}$ and $\widehat{b}_{A}$, 
so that $\text{rel}_l$ measures the variation in the decoder outputs attributable to axis $l$ (in the predictor setting, the treatment loading $\widehat{b}_{A,l}$ is absent). We rank the axes by $\text{rel}_l$ and retain the smallest set whose cumulative share of $\sum_l\text{rel}_l$ reaches a threshold $\gamma$ (here, 0.995). The number of retained axes is the selected dimension for that fit.

Since the ARD stage is run within each of the $K$ cross-fitting folds, folds may select different numbers of active axes. We set $d_z(\beta)$ to the maximum across folds, so that a single common dimension covers every fold, and then refit VAE-R at ($\beta$, $d_z(\beta)$) using the ordinary $\mathcal{N}(0,I)$ prior of Algorithms \ref{algo1} and \ref{algo2}. That is, the ARD fit itself is used only to determine $d_z(\beta)$. Since $\delta_Y$ and $\delta_A$ are linear in $Z_j$, redundant latent axes receive near-zero loadings and contribute little to the adjustments, so the procedure is robust to mild overspecification of $d_z$; see Section \ref{sec:simu} for more details. 

\subsubsection{KL weight} \label{sec:klw} For each $\beta \in \mathcal{B}$, the refit at ($\beta, d_z(\beta)$) yields out-of-fold CATE estimates and nuisances for all individuals. We use these outputs to compute a diagnostic that measures the cluster-level dependence remaining in the residuals after the latent adjustment. In the confounder setting, we use the cross-cluster covariance of the treatment residual and the treatment-effect-adjusted outcome residual, 
\begin{align}\label{eq:cc}
\text{CC}(\beta)=\frac 1J \sum_{j=1}^J \overline{R}^*_j \overline{V}^*_j, \quad  R^*_{ij}=\widehat{U}^*_{ij} - \widehat{\tau}(X_{ij}, W_j)\widehat{V}^*_{ij} \ ,
\end{align}
where $\overline{R}^*_j$ and $\overline{V}^*_j$ are cluster means. Subtracting $\widehat{\tau}\widehat{V}^*$ removes the treatment-effect signal in $\widehat{U}^*_{ij}$, so that $\text{CC}(\beta)$ isolates the remaining cluster-level confounding. When the latent adjustment captures $Z_j$, no shared cluster-level structure should remain in the residual products and $\text{CC}(\beta)$ should be close to zero; a nonzero value signals under- or over-shrinkage of the latent representation. In the predictor setting, the treatment residual carries no information about $Z_j$ by design, so we instead use the one-way intraclass correlation coefficient (ICC) of the same adjusted residual $R^*_{ij}$ (with $\widehat{V}_{ij}$ in place of $\widehat{V}^*_{ij})$, based on the ANOVA estimator with the standard correction for unequal cluster sizes \parencite{donner1986review}:
\begin{align}\label{eq:icc}
ICC(\beta) = \frac{\text{MS}_b - \text{MS}_w}{\text{MS}_b + (k_0 - 1) \text{MS}_w}, \quad k_0 = \frac{1}{J - 1} \left( n - \frac{\sum_{j=1}^{J} n_j^2}{n} \right) \ ,
\end{align}
where $\text{MS}_b$ and $\text{MS}_w$ are the between- and within-cluster mean squares of $R^*_{ij}$, with $J-1$ and $n-J$ degrees of freedom respectively, and $k_0$ is the effective cluster size. A well-chosen $\beta$ should leave no residual clustering in the outcome, so $ICC(\beta)$ should be close to zero. In both settings, we select
\[
\beta^* = \mathop{\text{argmin}}\limits_{\beta \in \mathcal{B}} \big| \Delta(\beta) \big|,
\]
where $\Delta(\beta)$ is $\text{CC}(\beta)$ in \eqref{eq:cc} for observational studies and $\text{ICC}(\beta)$ in \eqref{eq:icc}  for randomized trials, and report the estimates from the refit at ($\beta^*, d_z(\beta^*)$).

\subsubsection{Implementation} Algorithm \ref{algo3} summarizes our procedure for selecting the KL weight $\beta$ and the latent dimension size $d_z$. In our simulation study and applications below, we use $\mathcal{B}=\{1,2,4,6,8\}$, $d_{\max}=5$, $\rho_\alpha = 0.2$, $\gamma = 0.995$, $K = 5$, $K_{\rm inner} = 2$, $R = 300$ epochs (or above), a warmup fraction $\rho = 0.3$, and Adam with learning rate $5\times 10^{-3}$. The nuisance models and the CATE learner are specified in each application.

\begin{algorithm}[H]
\caption{\small{Data-Driven Selection of $(\beta, d_z)$ for VAE-R}}\label{algo3}
\begin{algorithmic}[1]
\small
\Require Data $\{Y_{ij}, T_{ij}, X_{ij}, W_j\}$, KL grid $\mathcal{B}$, maximum latent dimension $d_{\max}$, ARD subset fraction $\rho_\alpha$, relevance threshold $\gamma$, number of folds $K$, inner folds $K_{\rm inner}$.
\For{each $\beta \in \mathcal{B}$}
    \State \textbf{ARD stage:} Run Algorithm \ref{algo1} (or \ref{algo2}) with $d_z = d_{\max}$, replacing the $\mathcal{N}(0,I)$ prior in the VAE training step with the ARD prior with axis-specific variances, updated each epoch from a random fraction $\rho_\alpha$ of the training clusters. In each fold $k$, compute $\text{rel}_l$ for $l=1,\ldots,d_{\max}$ and record $d_z^{(k)}(\beta)$, the smallest number of top-ranked axes whose cumulative relevance share is at least $\gamma$ (here, 0.995).
    \State Set $d_z(\beta) = \max_k d_z^{(k)}(\beta)$.
    \State \textbf{Refit stage:} Run Algorithm \ref{algo1} (or \ref{algo2}) with $(\beta, d_z(\beta))$ and the $\mathcal{N}(0,I)$ prior to obtain out-of-fold $\hat\tau$, $\widehat{m}^*$, and $\widehat{e}^*$ (or $\widehat{e}$).
    \State Compute $\Delta(\beta) = \text{CC}(\beta)$ (confounder setting) or $\text{ICC}(\beta)$ (predictor setting) from the out-of-fold adjusted residuals.
\EndFor
\State Select $\beta^* = \operatorname{argmin}_{\beta \in \mathcal{B}} |\Delta(\beta)|$.
\Ensure CATE and ATE estimates from the refit at $(\beta^*, d_z(\beta^*))$.
\end{algorithmic}
\end{algorithm} \raggedbottom

\section{Simulation Study}\label{sec:simu}

\subsection{Simulation Designs and Evaluation} \label{sec:sim_design}

We conducted an extensive simulation study to assess the performance of VAE-R in the presence of unmeasured cluster-level confounders and predictors. We compare our proposed VAE-R with (1) an oracle method that treats the unmeasured cluster-level covariates as observed, (2) an ordinary R-learner that uses the measured covariates only, and (3) an R-learner with parametric fixed-effect nuisance estimation that regresses the outcome (or treatment) on the individual-level measured covariates and cluster dummies. Our simulation study consists of two main designs: one with unmeasured cluster-level confounder(s), and the other with unmeasured cluster-level predictor(s). Within each design, we vary the number of unmeasured confounders/predictors (i.e., 1 or 2) and their distributions (e.g., normal, exponential, binary). More specifically, for each simulated dataset, we generate 500 clusters, with cluster sizes drawn from a discrete uniform distribution on $[25, 35]$. The simulated data include 20 individual-level covariates $X_{ij}$, 10 cluster-level covariates $W_j$, and one or two unobserved cluster-level covariates $Z_j$. We generate the treatment variable from a logistic model and the outcome from a linear model, where we add higher-order terms in Section \ref{sec:simu_com}.
In the observational data setting, we include $Z_j$ in both the treatment and outcome models, whereas in the randomized trial setting, we include $Z_j$ only in the outcome model.

Throughout the simulation, we estimate the CATE $\tau(X_{ij}, W_j)$ and the ATE $\tau$ using Algorithms \ref{algo1} and \ref{algo2}, where the nuisance models are estimated via Elastic Net (or random forests in Section \ref{sec:simu_com}) and the CATE is estimated by weighted linear regression (or random forests in Section \ref{sec:simu_com}). We also vary the number of latent dimensions in the VAE, either fixing it at prespecified values or choosing it in the data-driven way described in Section \ref{sec:hyper}. We repeat each simulation condition 500 times ($m=1, \ldots, 500$) and measure the bias and the mean squared error (MSE), i.e.,
\begin{align*}
\text{Bias}_{\theta} &=\frac{1}{500}  \sum_{m=1}^{500} (\widehat{\theta}^{(m)}-\theta), \quad{}
\text{MSE}_{\theta} = \frac{1}{500} \sum_{m=1}^{500} (\widehat{\theta}^{(m)}-\theta)^2 .
\end{align*}

\noindent Here, $\theta$ denotes either $\tau(X_{ij}, W_j)$ or $\tau$, and $\widehat{\theta}^{(m)}$ is the estimate of $\theta$ from the $m$-th replication; for the CATE, the errors and squared errors are first averaged over all individuals within each replication. We present the results for the CATE estimates $\widehat{\tau}(X_{ij}, W_j)$ and the AIPW-based ATE estimates $\widehat{\tau}_{\rm ATE, AIPW}$ in the main text and report the results for the sample ATE $\widehat{\tau}_{\rm ATE}$ in Supplemental Appendix \ref{app:simu1}. \textsf{Python} code for our simulation study is available at the first author's GitHub repository.

We make two remarks. First, we report the performance of the data-driven selection of $\beta^*$ described in Section \ref{sec:hyper}, comparing the selected-$\beta^*$ results against those for each fixed $\beta$ in the grid and against the oracle in terms of bias or MSE. The bias or MSE of VAE-R with the selected $\beta^*$ is comparable to that of the oracle in each setting; see Supplemental Appendix \ref{app:simu0} for more details. Second, we conducted an additional simulation where $Z_j$ moderates the treatment effect while VAE-R is fitted as is, without accommodating effect modification. In the confounder setting, the CATE estimates are biased for the $Z$-marginal CATE, whereas the AIPW-based ATE estimates show negligible bias. In the predictor setting, both the CATE and the AIPW-based ATE estimates show almost no bias, with somewhat larger MSE than in the absence of moderation; see Supplemental Appendix \ref{app:simu_mod} for more details.

\subsection{Results with Unmeasured Cluster-Level  Confounders}

Table \ref{tab:simu_D1} summarizes the simulation results for the CATE and ATE estimates obtained by the oracle, ordinary, and proposed VAE-R methods in the presence of unmeasured cluster-level  confounders. We implement our proposed method either with a fixed number of latent dimensions (i.e., $d_z = c$ for an integer $c$) or with the data-driven number chosen by the ARD strategy. Here, we report the average of the chosen latent dimension size, $\bar d_{z,ARD}$, across the 500 replications. In the first condition, which contains one normally distributed $Z_{1j}$, $\bar d_{z,ARD}$ (1.13) is close to the true value of 1, and our proposed method, both with the true latent dimension ($d_z=1$) and with the ARD-based latent dimension, shows near-zero bias and MSE comparable to that of the oracle method for both the CATE and ATE estimates. In contrast, the ordinary method shows substantial bias and MSE.

\begin{table}[h!]
    \centering
 \begin{threeparttable}
    \caption{Simulation results with unmeasured cluster-level confounders \label{tab:simu_D1}}
\small
    \begin{tabular}{lC{1.5cm}C{1.5cm}|C{1.5cm}C{1.5cm}}
\hline
& \multicolumn{2}{c|}{CATE} & \multicolumn{2}{c}{ATE$_{\rm AIPW}$} \\
& Bias & MSE & Bias & MSE \\
\hline
\textit{Normal $Z_{1j}$} &  &  &  &  \\
$\ $ Oracle & -0.001 & 0.013 & -0.001 & 0.000 \\
$\ $ Ordinary & 1.364 & 1.991 & 1.422 & 2.031 \\
$\ $ Ours ($d_z=1$) & -0.007 & 0.015 & -0.000 & 0.000 \\
$\ $ Ours ($\bar d_{z,ARD}=1.13$) & -0.005 & 0.015 & 0.001 & 0.000 \\
\hline
\textit{Exponential $Z_{1j}$} &  &  &  &  \\
$\ $ Oracle & -0.001 & 0.013 & -0.000 & 0.000 \\
$\ $ Ordinary & 1.233 & 1.719 & 1.286 & 1.656 \\
$\ $ Ours ($d_z=1$) & -0.007 & 0.015 & 0.009 & 0.001 \\
$\ $ Ours ($d_z=2$) & -0.003 & 0.014 & 0.011 & 0.001 \\
$\ $ Ours ($\bar d_{z,ARD}=1.15$) & -0.005 & 0.014 & 0.009 & 0.001 \\
\hline
\textit{Binary $Z_{1j}$} &  &  &  &  \\
$\ $ Oracle & 0.000 & 0.012 & -0.000 & 0.000 \\
$\ $ Ordinary & 1.483 & 2.250 & 1.546 & 2.392 \\
$\ $ Ours ($d_z=1$) & -0.005 & 0.017 & -0.004 & 0.001 \\
$\ $ Ours ($d_z=2$) & -0.002 & 0.015 & -0.002 & 0.000 \\
$\ $ Ours ($\bar d_{z,ARD}=1.08$) & -0.004 & 0.016 & -0.003 & 0.001 \\
\hline
\textit{Normal $Z_{1j}, Z_{2j}$} &  &  &  &  \\
$\ $ Oracle & -0.001 & 0.014 & -0.002 & 0.000 \\
$\ $ Ordinary & 0.565 & 0.405 & 0.589 & 0.352 \\
$\ $ Ours ($d_z=1$) & -0.110 & 0.038 & -0.106 & 0.012 \\
$\ $ Ours ($d_z=2$) & -0.050 & 0.019 & -0.050 & 0.003 \\
$\ $ Ours ($d_z=3$) & -0.023 & 0.016 & -0.023 & 0.001 \\
$\ $ Ours ($d_z=5$) & -0.008 & 0.014 & -0.007 & 0.000 \\
$\ $ Ours ($\bar d_{z,ARD}=3.36$) & -0.008 & 0.014 & -0.008 & 0.001 \\
\hline
\textit{Normal $Z_{1j}$ and Binary $Z_{2j}$} &  &  &  &  \\
$\ $ Oracle & -0.001 & 0.014 & -0.000 & 0.000 \\
$\ $ Ordinary & 0.545 & 0.377 & 0.568 & 0.327 \\
$\ $ Ours ($d_z=1$) & -0.122 & 0.037 & -0.116 & 0.014 \\
$\ $ Ours ($d_z=2$) & -0.051 & 0.019 & -0.048 & 0.003 \\
$\ $ Ours ($d_z=3$) & -0.020 & 0.016 & -0.018 & 0.001 \\
$\ $ Ours ($d_z=5$) & -0.007 & 0.014 & -0.004 & 0.000 \\
$\ $ Ours ($\bar d_{z,ARD}=3.33$) & -0.008 & 0.014 & -0.005 & 0.000 \\
\hline
    \end{tabular}
\begin{flushleft}
\footnotesize  NOTE: MSE = mean squared error; $d_z$ = the number of latent variables; $\bar d_{z,ARD}$ = the average number of latent variables selected by the ARD strategy across 500 replications.
\end{flushleft}
 \end{threeparttable}
\end{table}

When $Z_{1j}$ is exponentially or binary distributed (i.e., the \textit{Exponential $Z_{1j}$} and \textit{Binary $Z_{1j}$} conditions), we vary the number of latent dimensions by fixing it at 1 or 2 or selecting it using the ARD strategy. Again, $\bar d_{z,ARD}$ is close to the true value of 1 in both conditions (1.15 and 1.08, respectively). Moreover, although our proposed method still uses a normal prior, it shows robust performance across the different dimension choices, with negligible bias and oracle-equivalent MSE for both the CATE and ATE estimates. Not surprisingly, the ordinary method shows the worst performance.

In the presence of two unmeasured cluster-level  confounders (i.e., the \textit{Normal $Z_{1j}, Z_{2j}$} and \textit{Normal $Z_{1j}$ and Binary $Z_{2j}$} conditions), we vary the number of latent dimensions with fixed values of 1, 2, 3, and 5, as well as the ARD-based data-driven value. We observe that $\bar d_{z,ARD}$ (3.36 and 3.33) is closer to 3, one larger than the true number of unmeasured confounders. Although 2 is the true dimension in these conditions, our proposed method with $d_z = 1$ or $2$ shows some bias and slightly larger MSE for the CATE and ATE estimates compared to the oracle. This may be because the KL regularization shrinks each latent axis toward zero, so that additional axes help recover the full cluster-level shifts. However, with $d_z = 3$ or $5$, or with the ARD-based latent dimension, our proposed method shows much smaller bias and oracle-equivalent MSE. Overall, VAE-R,  with the data-driven ARD strategy, performs well across the simulation conditions.
 
\subsection{Results with Unmeasured Cluster-Level  Predictors}

Table \ref{tab:simu_D2} summarizes the simulation results for the CATE and ATE estimates obtained by the oracle, ordinary, and proposed VAE-R methods in the presence of unmeasured cluster-level predictors. As in Table \ref{tab:simu_D1}, we implement our proposed method either with a fixed number of latent dimensions or with the data-driven ARD strategy, reporting the average chosen latent dimension size $\bar d_{z,ARD}$. In this design, all methods are expected to be unbiased, because unmeasured predictors affect only the precision of the target estimates. In the first condition, which contains one normally distributed $Z_{1j}$, $\bar d_{z,ARD}$ (1.37) is close to the true value of 1. Our proposed method, both with the true latent dimension ($d_z=1$) and with the ARD-based latent dimension, shows negligible bias and provides efficiency gains, with MSE comparable to that of the oracle method. The ordinary method shows the largest MSE, as expected.

\begin{table}[h!]
    \centering
 \begin{threeparttable}
    \caption{Simulation results with unmeasured cluster-level predictors \label{tab:simu_D2}}
\small
    \begin{tabular}{lC{1.5cm}C{1.5cm}|C{1.5cm}C{1.5cm}}
\hline
& \multicolumn{2}{c|}{CATE} & \multicolumn{2}{c}{ATE$_{\rm AIPW}$} \\
& Bias & MSE & Bias & MSE \\
\hline
\textit{Normal $Z_{1j}$} &  &  &  &  \\
$\ $ Oracle & -0.000 & 0.011 & -0.000 & 0.000 \\
$\ $ Ordinary & 0.000 & 0.054 & 0.000 & 0.001 \\
$\ $ Ours ($d_z=1$) & -0.000 & 0.012 & 0.000 & 0.000 \\
$\ $ Ours ($\bar d_{z,ARD}=1.37$) & -0.000 & 0.012 & 0.000 & 0.000 \\
\hline
\textit{Normal $Z_{1j}, Z_{2j}$} &  &  &  &  \\
$\ $ Oracle & 0.000 & 0.011 & 0.000 & 0.000 \\
$\ $ Ordinary & 0.001 & 0.066 & 0.002 & 0.002 \\
$\ $ Ours ($d_z=1$) & 0.001 & 0.012 & 0.000 & 0.000 \\
$\ $ Ours ($d_z=2$) & 0.001 & 0.012 & 0.001 & 0.000 \\
$\ $ Ours ($d_z=3$) & 0.001 & 0.012 & 0.001 & 0.000 \\
$\ $ Ours ($d_z=5$) & 0.001 & 0.012 & 0.001 & 0.000 \\
$\ $ Ours ($\bar d_{z,ARD}=1.64$) & 0.001 & 0.012 & 0.000 & 0.000 \\
\hline
    \end{tabular}
\begin{flushleft}
\footnotesize  NOTE: MSE = mean squared error; $d_z$ = the number of latent variables; $\bar d_{z,ARD}$ = the average number of latent variables selected by the ARD strategy across 500 replications.
\end{flushleft}
 \end{threeparttable}
\end{table}

When there are two normally distributed unmeasured cluster-level predictors, we again vary the number of latent dimensions with fixed values of 1, 2, 3, and 5, as well as the ARD-based data-driven value. We observe that $\bar d_{z,ARD}$ is 1.64, below the true value of 2. This is expected because in the predictor setting, the two predictors affect the data only through the single outcome shift $\delta_Y(Z_j)$, so one latent dimension suffices to capture it, which also explains why all fixed choices of $d_z$ perform equally well. Our proposed method, with both the ARD-based value and the fixed numbers of latent dimensions, shows no bias and MSE comparable to that of the oracle method. Overall, VAE-R provides efficiency gains in multisite randomized trials across the simulation conditions.

\subsection{Results Compared with Fixed Effects Models}\label{sec:simu_com}

We compare the performance of our VAE-R approach with the R learner that uses parametric fixed effects nuisance estimation. In this design, we increase the complexity of the nuisance models by adding higher-order terms in the outcome and treatment models. Our VAE-R approach uses random forests to fit the two nuisance models and the CATE model. In contrast, the fixed effects approach fits generalized linear regressions that include cluster dummies as additional predictors for the nuisance models (the outcome model only in the predictor setting, where the propensity score is known), and a weighted linear regression for the CATE model. Therefore, the fixed effects approach requires the correct functional forms of the nuisance models, as well as the CATE model. 

To disentangle the features of the fixed effects approach, we implement it with three variants: (i) with cross-fitting and the correct functional form,  denoted as FE (CF, correct), where the average of the cluster-specific intercepts estimated on the training clusters is used for held-out clusters; (ii) without cross-fitting and with the correct functional form, denoted as FE (no CF, correct); and (iii) without cross-fitting and with main-effect terms only, denoted as FE (no CF, incorrect). The ``correct'' functional form includes the true higher-order terms of the data-generating model, including interaction, squared, and hinge terms; variant (ii) is thus an infeasible ceiling rather than a practical competitor, whereas variant (iii) reflects what an analyst would fit in practice. 

Table \ref{tab:simu_D3} summarizes the results in the nonlinear setting. In the confounder setting, VAE-R yields small bias and substantially lower MSE than FE (CF, correct) and FE (no CF, incorrect) for both the CATE and ATE estimates, and in the predictor setting, it reduces the MSE by increasing precision. In contrast, FE (CF, correct) performs poorly even with the correct functional form. This is because a fixed effects model cannot supply intercepts for new clusters, so using the grand-mean intercept for them leaves the cluster-level confounding or predictor variation unadjusted. In contrast, FE (no CF, correct) attains near-oracle performance, as expected for a correctly specified parametric model, but it requires knowledge of the correct functional form, which is unavailable in practice. FE (no CF, incorrect) shows bias in the confounder setting where both the outcome and the treatment models are misspecified. In the predictor setting, it shows negligible bias because the known propensity score protects the R-learner and the AIPW estimator against outcome model misspecification, but the MSE of CATE is large because the linear CATE model omits the higher-order terms. Overall, among the methods, VAE-R is the only approach that combines flexible machine learning, cluster-level adjustment, and valid cross-fitting, and it performs closest to the infeasible benchmarks, i.e., oracle and FE (no CF, correct).

\begin{table}[h!]
    \centering
 \begin{threeparttable}    
    \caption{Simulation results compared with fixed effect models\label{tab:simu_D3}} 
\small
    \begin{tabular}{l|l|C{1.5cm}C{1.5cm}|C{1.5cm}C{1.5cm}}
\hline
\multirow{2}{*}{Design} & \multirow{2}{*}{Condition}  & \multicolumn{2}{c|}{CATE} & \multicolumn{2}{c}{ATE$_{\rm AIPW}$} \\
& & Bias & MSE & Bias & MSE \\
\hline
\multirow{7}{*}{Confounders} & \textit{Normal $Z_{1j}$} &  &  &  &  \\
& $\ $ Oracle & -0.001 & 0.013 & -0.001 & 0.000 \\
& $\ $ Ours ($\bar d_{z,ARD}=1.04$) & 0.011 & 0.037 & 0.008 & 0.001 \\
& $\ $ FE (CF, correct) & 0.468 & 0.257 & 0.469 & 0.222 \\
& $\ $ FE (no CF, correct) & -0.001 & 0.011 & -0.001 & 0.000 \\
& $\ $ FE (no CF, incorrect) & 0.115 & 0.127 & 0.115 & 0.025 \\
\hline
\multirow{7}{*}{Predictors} & \textit{Normal $Z_{1j}$} &  &  &  &  \\
& $\ $ Oracle & -0.000 & 0.011 & -0.000 & 0.000 \\
& $\ $ Ours ($\bar d_{z,ARD}=1.01$) & -0.000 & 0.068 & 0.000 & 0.000 \\
& $\ $ FE (CF, correct) & 0.000 & 0.399 & 0.001 & 0.009 \\
& $\ $ FE (no CF, correct) & -0.000 & 0.009 & -0.000 & 0.000 \\
& $\ $ FE (no CF, incorrect) & -0.000 & 0.373 & -0.000 & 0.000 \\
\hline
    \end{tabular}
\begin{flushleft}
\footnotesize  NOTE: MSE = mean squared error; $\bar d_{z,ARD}$ = the average number of latent variables selected by the ARD strategy across 500 replications; FE = fixed effects models; CF = cross-fitting.
\end{flushleft} 
 \end{threeparttable}
\end{table}

\section{Empirical Examples}

\subsection{HSLS Data and Results}

We use the HSLS dataset to study the effect of taking an advanced math course in ninth grade on students' ninth-grade math achievement, following \textcite{pan2026learning}. Students are nested within schools, and the analytic sample consists of $n=17,121$ students in $J=854$ schools. The outcome $Y_{ij}$ is the student’s weighted GPA in mathematics in ninth grade, which adjusts for the difficulty level of the courses taken. The treatment $T_{ij}$ indicates whether the student took an advanced math course (geometry or higher) during the 2009-10 academic year, with $T_{ij} = 1$ for advanced course-taking and $T_{ij} = 0$ otherwise. The pre-treatment covariates include 13 student-level covariates, such as demographic characteristics (e.g., gender and race/ethnicity), prior math achievement, math identity, math self-efficacy, educational expectations, household socioeconomic status, and parental education, as well as 10 school-level covariates, such as the student-to-math-teacher ratio, math course requirements, and school type. Missing covariate values are imputed using predictive mean matching.

We apply the proposed VAE-R method for unmeasured cluster-level confounders (Algorithm \ref{algo1}), with the KL weight and latent dimension selected as described in Section \ref{sec:hyper}; the selected values are $\beta^*=1$ and $d_z=2$. The nuisance models and the CATE model are estimated by random forests, with $K=5$ cluster-level folds. For comparison, we also apply the ordinary R-learner using the measured covariates only, as in the simulation study. Standard errors are obtained from 500 cluster bootstrap replications. \textsf{Python} code for our real data applications is available at the first author's GitHub repository.

\begin{figure}[b!]
    \centering
\subcaptionbox{CATE distributions \label{fig:hsls_CATE}}[.4\textwidth]{
    \includegraphics[width=1\linewidth]{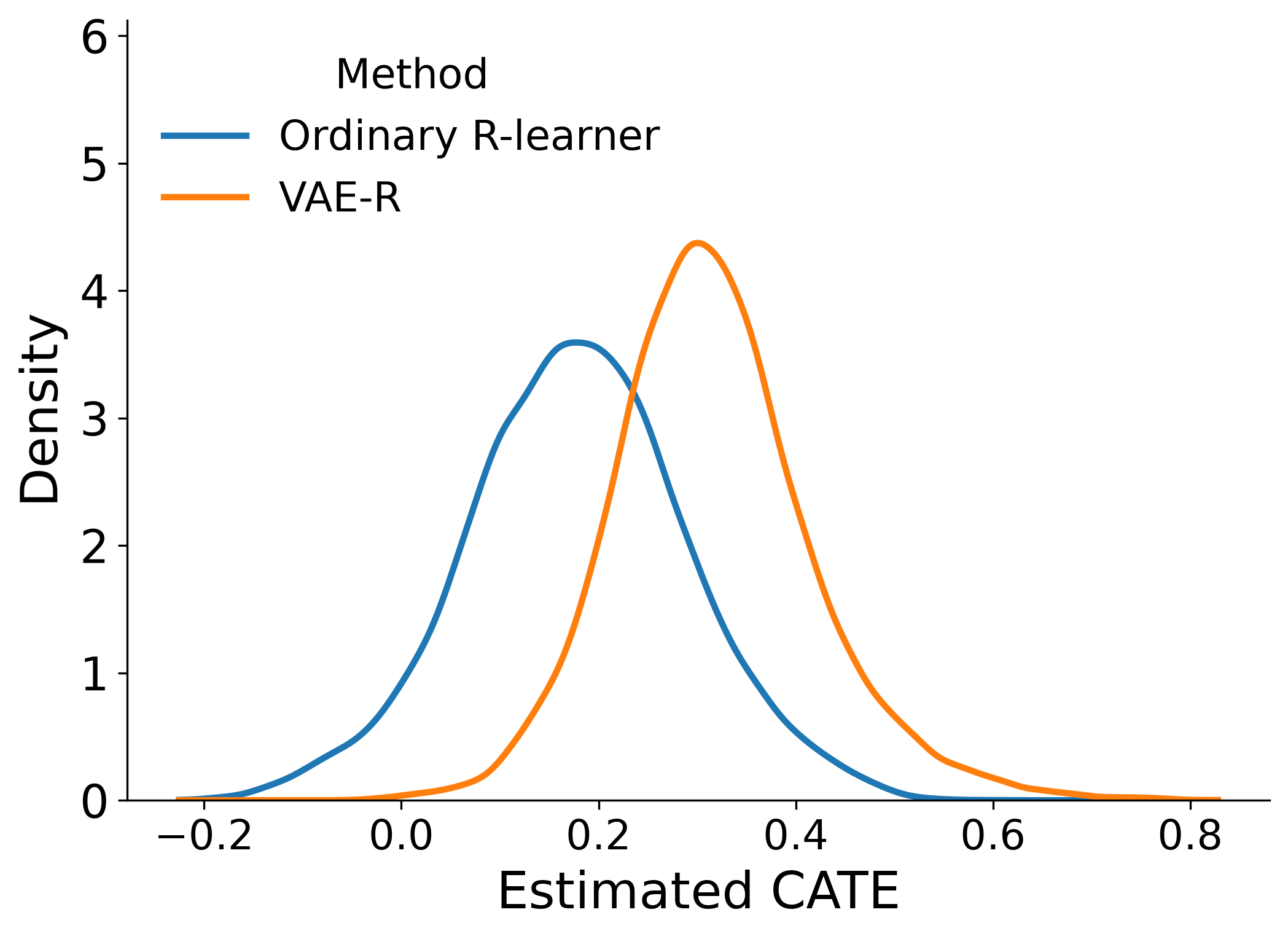}}
\subcaptionbox{ATE estimates\label{fig:hsls_ATE}}[.5\textwidth]{
    \includegraphics[width=1\linewidth]{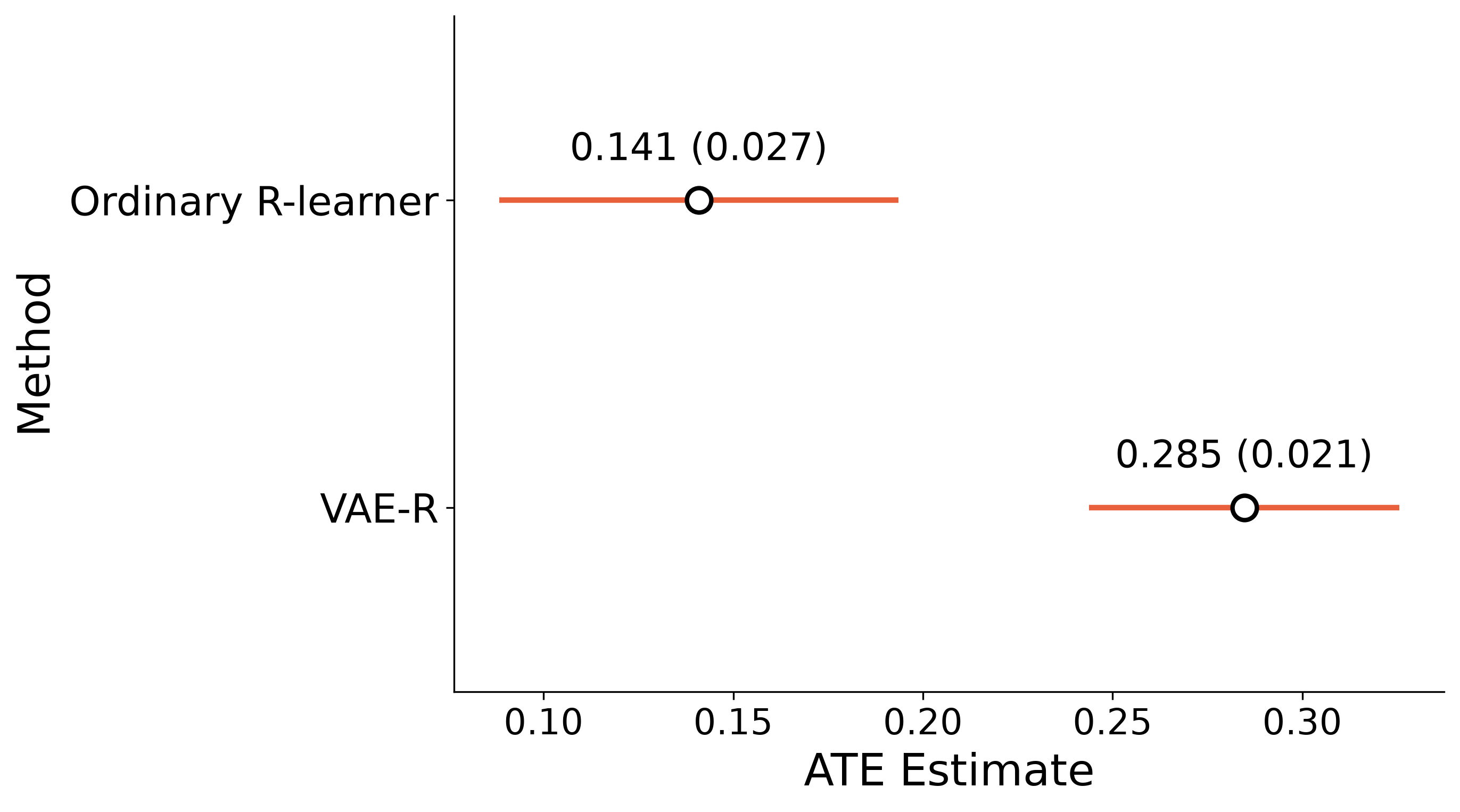}}
    \caption{Distributions of the estimated conditional average treatment effect (CATE), and average treatment effect (ATE) estimates with 95\% cluster bootstrap confidence intervals, for the ordinary R-learner using measured covariates only and the proposed VAE-R, in the High School Longitudinal Study (HSLS) data.}\label{fig:hsls}
\end{figure}

Figure \ref{fig:hsls} summarizes the CATE and ATE estimates. The distribution of CATE estimates from VAE-R is shifted in a positive direction, relative to that from the ordinary R-learner. The AIPW-based ATE estimate from VAE-R is 0.285 (bootstrap SE = 0.021), compared with 0.141 (SE = 0.027) from the ordinary R-learner, and the two 95\% confidence intervals do not overlap. That is, VAE-R yields a significantly larger estimated effect of advanced course-taking. Although the true effects are unknown in empirical data, this difference indicates that the ordinary R-learner and VAE-R lead to substantively different conclusions and supports the presence of cluster-level unmeasured confounding in the HSLS data.

\subsection{NSLM Data and Results}

We use the NSLM dataset from a multisite randomized trial to study the effect of a growth mindset intervention on students' academic achievement in high school. The sample contains $n=9,590$ students from $J=62$ schools. The outcome $Y_{ij}$ is each student's post-intervention English GPA. The treatment $T_{ij}$ indicates whether the student was randomly assigned to the growth mindset intervention or a control condition, each delivered through brief online sessions. Assignment was made by the computer software with probability 0.5 for every student, so treatment was randomized within each school. There are 7 student-level covariates, including gender, race/ethnicity, special education status, free or reduced-price-lunch eligibility, English language learner status, freshman status, and expectations for success in math, and 3 school-level covariates, including the percentage of students receiving free or reduced-price lunch, poverty percentage, and minority percentage. Missing covariates are imputed by predictive mean matching.

We apply VAE-R for unmeasured cluster-level predictors (Algorithm \ref{algo2}) with the selected KL weight $\beta^*=2$ and latent dimension $d_z=1$. The treatment probability is fixed at 0.5 by design. The outcome and CATE models are estimated by random forests, with $K=5$ cluster-level folds. Standard errors are obtained from 500 cluster bootstrap replications.

\begin{figure}[!ht]
    \centering
\subcaptionbox{CATE distributions \label{fig:nslm_CATE}}[.4\textwidth]{
    \includegraphics[width=1\linewidth]{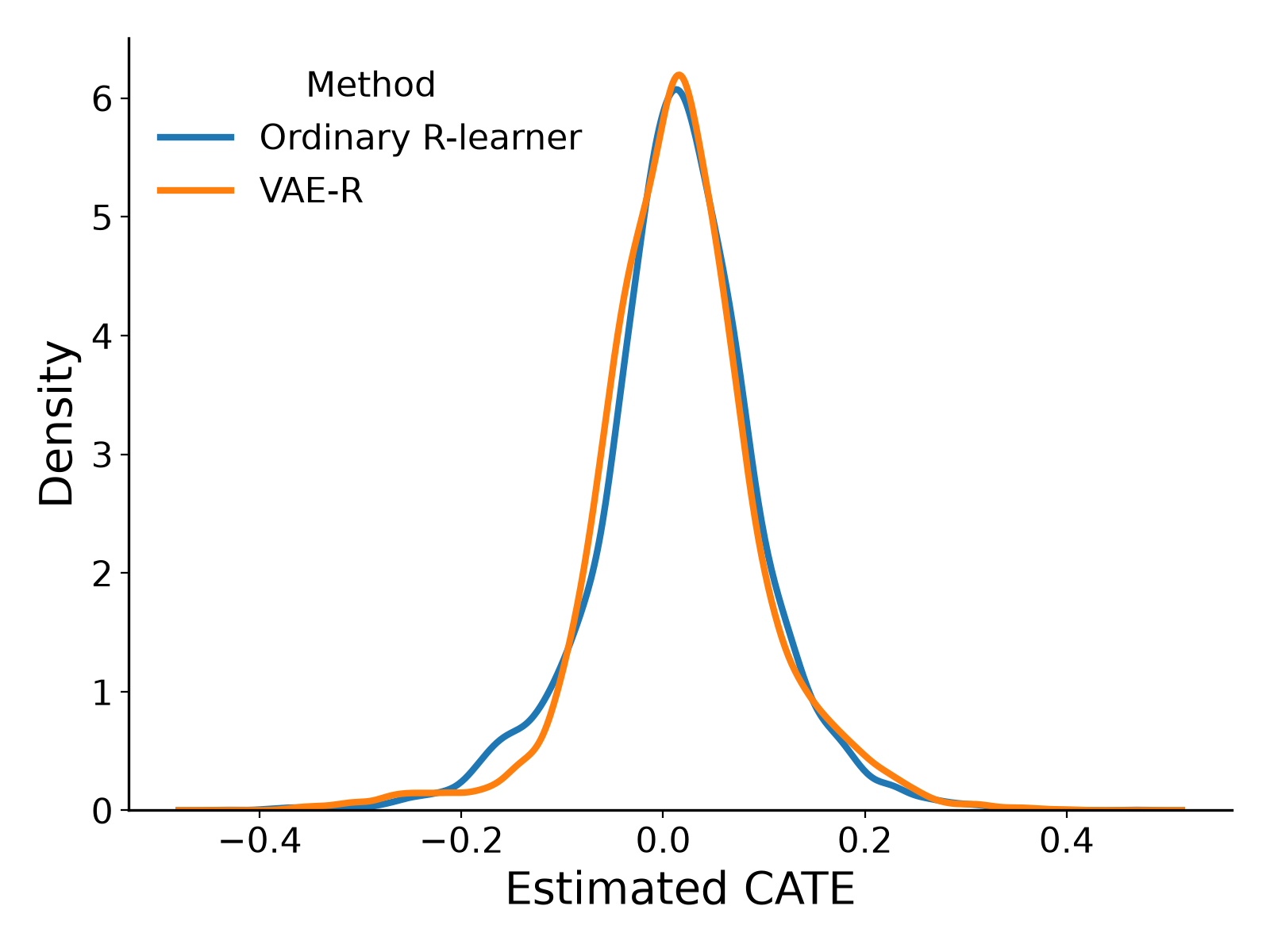}}
\subcaptionbox{ATE estimates\label{fig:nslm_ATE}}[.5\textwidth]{
    \includegraphics[width=1\linewidth]{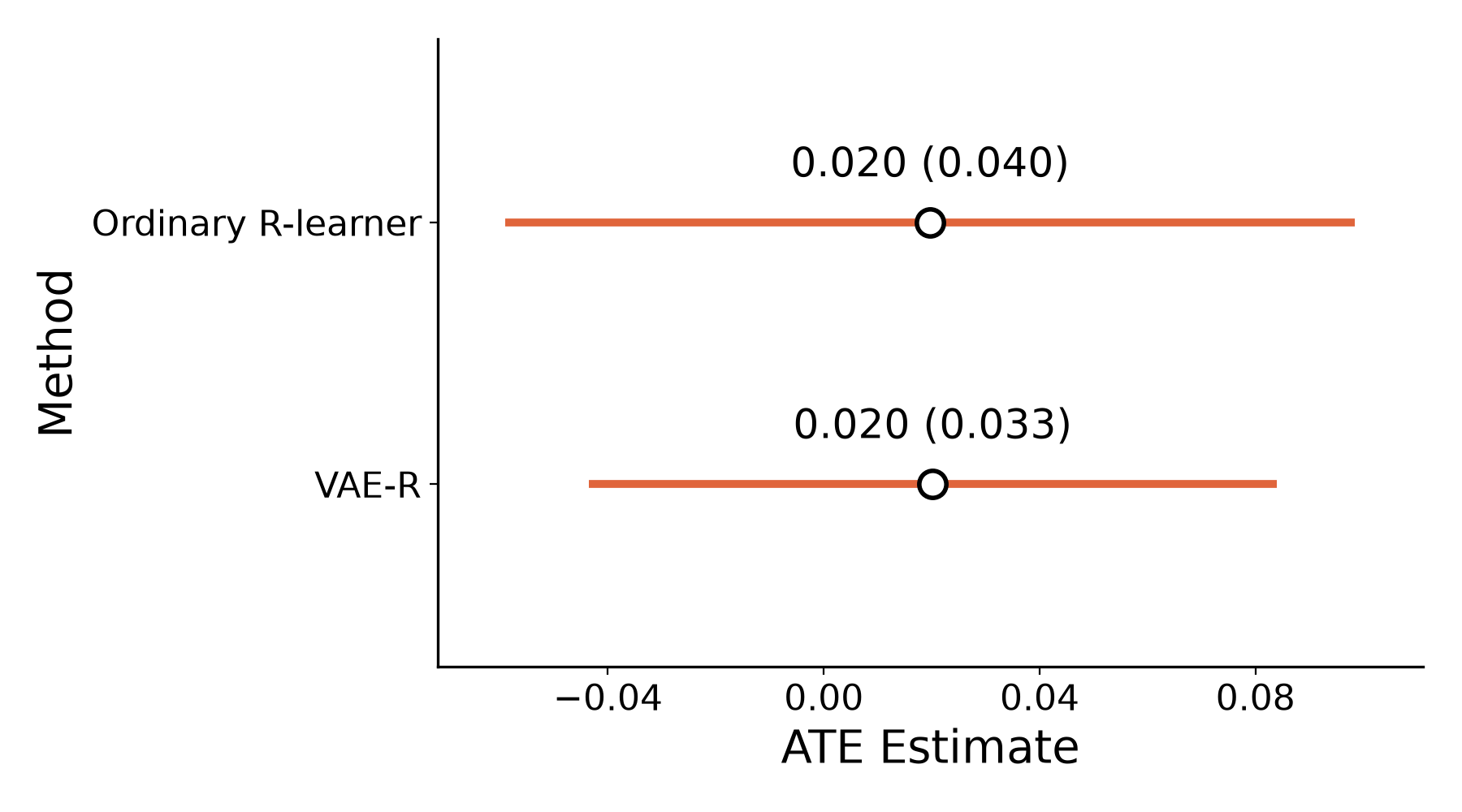}}
    \caption{Distributions of the estimated conditional average treatment effect (CATE), and average treatment effect (ATE) estimates with 95\% cluster bootstrap confidence intervals, for the ordinary R-learner using measured covariates only and the proposed VAE-R, in the National Study of Learning Mindsets (NSLM) data.}\label{fig:nslm}
\end{figure}

Figure \ref{fig:nslm} summarizes the results. The estimated CATE distribution from VAE-R has a slightly higher peak and narrower tails than that from the ordinary R-learner. The AIPW-based ATE estimate is 0.020 (bootstrap SE = 0.033) for VAE-R and 0.020 (SE = 0.040) for the ordinary R-learner. The point estimates are nearly identical, and VAE-R reduces the standard error by 18\%, though neither estimate significantly differs from zero. These findings suggest that VAE-R improves the precision of treatment effect estimation in the NSLM data by adjusting for unmeasured cluster-level predictors.

\section{Discussion and Conclusions}

We proposed the VAE-R approach for estimating the CATE and ATE in multilevel data with unmeasured cluster-level covariates, in both observational studies and randomized trials. This approach integrates ideas from latent DML and the R-learner by inferring cluster-level latent variables from outcome and treatment residuals via a VAE and using the resulting latent adjustments in the R-learner objective. In particular, we specify the decoder through an offset likelihood, so that the latent variables capture only the cluster-level variation left unexplained by the measured covariates, and we provide data-driven strategies for choosing the latent dimension and the KL weight. Our simulation study shows that the VAE-R substantially reduces the bias of the ordinary R-learner in multilevel observational data and improves its efficiency in multisite randomized trials. In the HSLS data, the ATE estimate obtained by VAE-R  differs substantially from that of the ordinary R-learner, whereas in the NSLM data, VAE-R reduces the bootstrap standard error of the ATE estimate by 18\%. 

To elaborate, VAE-R addresses cluster-level unmeasured confounding by learning latent cluster-level structure from the within-cluster dependence of the outcome and treatment residuals. Since the encoder is a shared, amortized mapping from a cluster's residual summaries to its latent representation, VAE-R can construct latent adjustments for clusters not used in training. This feature allows VAE-R to accommodate (cluster-level) cross-fitting, which has become standard for machine learning methods in causal inference \parencite{chernozhukov2018doubleml}. VAE-R has several advantages over existing approaches. As shown in Section \ref{sec:simu_com}, parametric fixed effects models require specifying the functional relationships between the measured covariates and the outcome and treatment, and cluster-specific fixed effects cannot be directly estimated for clusters not in the training sample. Likewise, parametric random effects models require specifying these functional relationships and additionally assume that the cluster effects are independent of treatment assignment, which is violated by definition when the unmeasured cluster-level covariate is a confounder. VAE-R instead uses residual dependence within clusters to learn latent adjustments without imposing a parametric model that links the latent cluster effect to observed cluster identifiers. Our VAE-R also differs from existing robust machine learning methods for multilevel observational data \parencite[e.g.,][]{suk2022robust, suk2023tuning} in that it explicitly learns latent cluster-level representations and uses them to construct the residual adjustments tailored to both observational studies and randomized trials.  

When implementing VAE-R, researchers may have concerns about the sign and rotational indeterminacy of latent variables extracted from a VAE. That is, the latent variables are identified only up to sign and, when $d_z \geq 2$, up to rotation. This indeterminacy is not consequential for causal effect estimation in our framework because the latent variables enter the estimation only through the decoder adjustments $\delta_Y(Z_j) = b_Y^\top Z_j$ and $\delta_A(Z_j) = b_A^\top Z_j$, and if the latent coordinates are reoriented and the decoder loadings are transformed accordingly, the resulting adjustments remain unchanged. Hence, the residuals and the causal effect estimates depend on the latent adjustments rather than on the particular orientation of the latent coordinates. Similarly, our simulations suggest that moderate overspecification of the latent dimension size in our approach is relatively benign because superfluous latent axes receive near-zero loadings and thus contribute little to the latent adjustments, as shown in Section \ref{sec:simu}.

Regarding efficiency gains, incorporating unmeasured cluster-level predictors can reduce unexplained between-cluster variation in the outcome. This parallels the broader principle of prognostic covariate adjustment in randomized trials and is related to prior work by \textcite{gagnon2023precise}, who use auxiliary information from external data to improve the precision of ATE estimation in randomized trials. In our framework, we recover this prognostic information from the within-cluster dependence of the outcome residuals without external data, and use the resulting latent adjustment to improve efficiency. This gain may be partly offset by the additional variability introduced by estimating the latent representation. Therefore, VAE-R is expected to reduce variance when the prognostic information from the latent variables exceeds the VAE-introduced estimation error.

This paper has some limitations that suggest directions for future research. First, we assume that the unmeasured cluster-level covariates shift the outcome additively and do not moderate the treatment effect. Supplemental Appendix \ref{app:simu_mod} shows that when this assumption is violated, the CATE estimates are biased in the confounder setting, whereas the AIPW-based ATE estimates show negligible bias. Future work could extend the decoder to accommodate interactions between the cluster-level latent variables and treatment. Second, we focus on unmeasured covariates that arise at the cluster level. Inferring unmeasured individual-level variables would generally require additional information or stronger assumptions, such as proxy variables or parametric latent variable models. Future research could extend VAE-R to individual-level latent variables under such conditions. Overall, VAE-R serves as a useful tool for estimating treatment effects in multilevel studies when cluster-level covariates are unmeasured.

\printbibliography

\newpage 

\setcounter{equation}{0}
\setcounter{figure}{0}
\setcounter{table}{0}
\setcounter{page}{1}
\setcounter{section}{0}
\renewcommand{\theequation}{S\arabic{equation}}
\renewcommand{\thefigure}{S\arabic{figure}}
\renewcommand{\thesection}{S\arabic{section}}
\renewcommand{\thetable}{S\arabic{table}}
\renewcommand{\thefigure}{S\arabic{figure}}
\setlength{\parindent}{30pt}

\setcounter{secnumdepth}{2}

\begin{center}
{\Large
\textbf{Supplementary Materials}
}
\end{center}

\section{The KL Weight Selection}\label{app:simu0}

We conduct simulation studies to evaluate the KL weight selection criteria described in Section \ref{sec:klw} under settings with  unmeasured cluster-level confounders or predictors. The simulation design follows Section \ref{sec:sim_design}. Across different distributions of unmeasured covariates, we compare VAE-R using each fixed KL weight in $\{1, 2, 4, 6, 8\}$ with VAE-R using a KL weight selected from the same grid by the proposed criteria. 

\begin{figure}[htbp]
    \centering
    \includegraphics[width=0.9\linewidth]{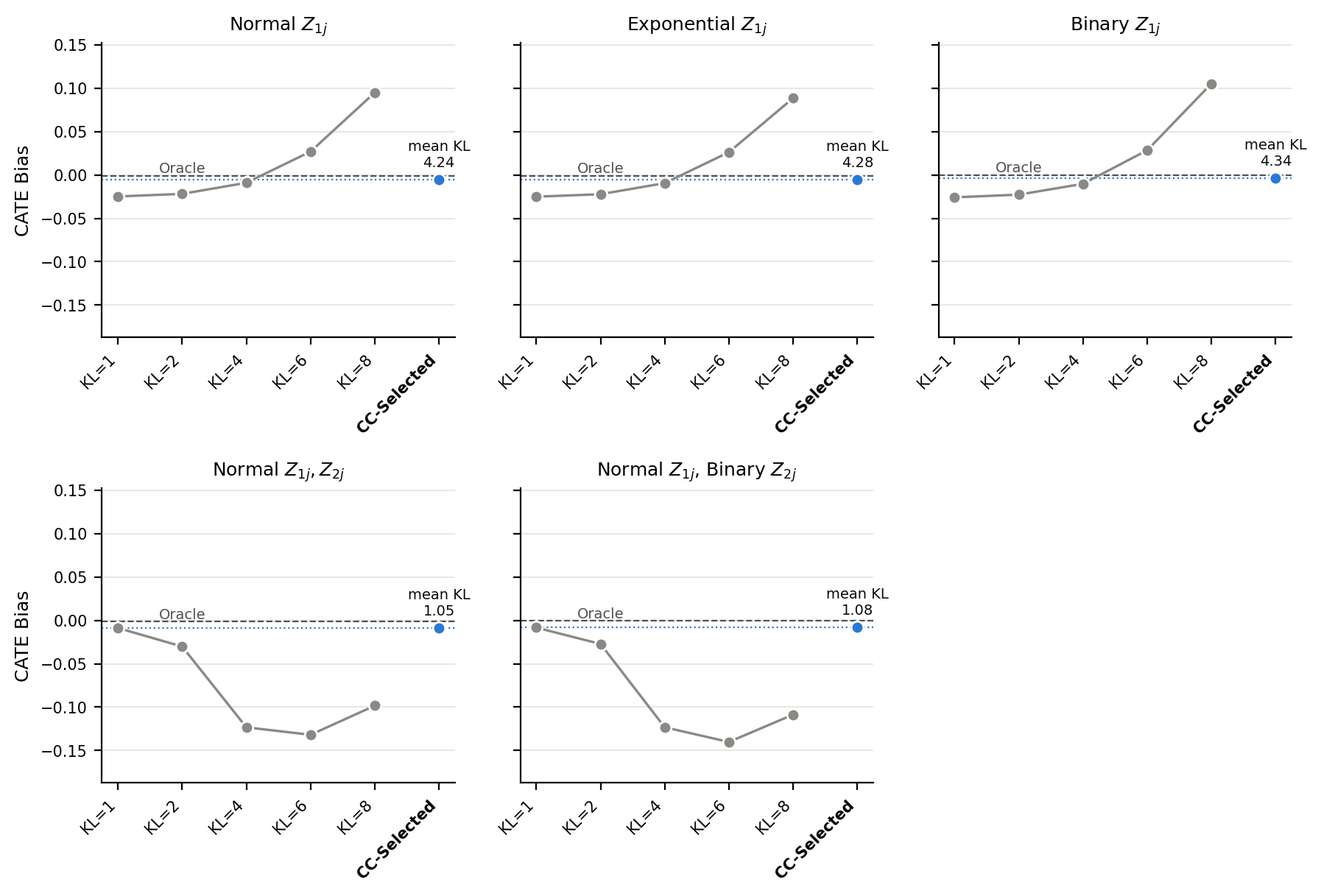}
    \caption{Bias of estimated conditional average treatment effect (CATE) under fixed KL weight and under the KL weight selected by the cross-cluster covariance (CC) criterion in the unmeasured confounder setting.}\label{fig:KLConf}
\end{figure}

Figure \ref{fig:KLConf} summarizes the results for the unmeasured confounder setting. Across all scenarios, the cross-cluster covariance criterion selects the KL weights that yield the smallest bias in CATE estimation among the candidate KL weights. VAE-R using the selected weight achieves performance comparable to that of the oracle model.

\begin{figure}[htbp]
    \centering
    \includegraphics[width=0.9\linewidth]{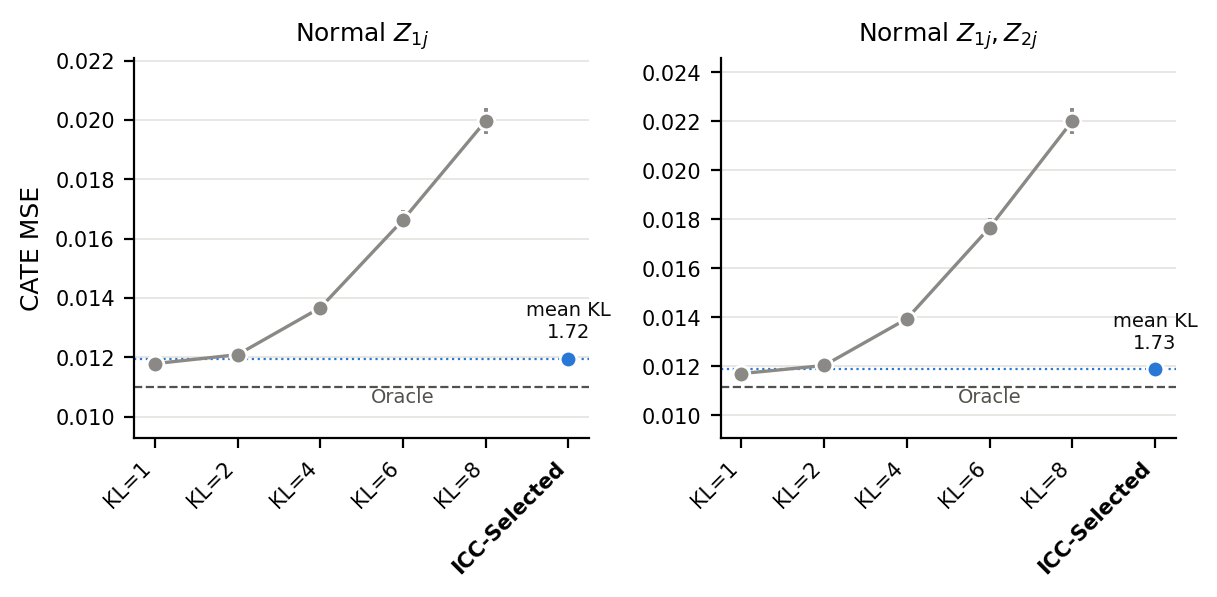}
    \caption{Mean squared error (MSE) of estimated conditional average treatment effect (CATE) under fixed KL weight and under the KL weight selected by the intraclass correlation coefficient (ICC) criterion in the unmeasured predictor setting.}\label{fig:KLPred}
\end{figure}

Figure \ref{fig:KLPred} summarizes the results for the unmeasured predictor setting. Since CATE estimates are approximately unbiased, we evaluate the performance using CATE MSE under this setting. The ICC criterion selects the KL weights with the lowest CATE MSE over the candidate grid. The VAE-R with the selected KL weights performs similarly to the oracle model.

\clearpage

\section{The Results of ATE: Sample Average}\label{app:simu1}

\begin{table}[h!]
    \centering
 \begin{threeparttable}
    \caption{ATE Estimates Obtained by the Sample Average in Confounder and Predictor Settings\label{app:tab:simu_naiveATE}}
\small
    \begin{tabular}{lC{1.5cm}C{1.5cm}|C{1.5cm}C{1.5cm}}
\hline
                & \multicolumn{2}{c|}{Confounder: ATE} & \multicolumn{2}{c}{Predictor: ATE} \\
                & Bias & MSE & Bias & MSE \\
    \hline
\textit{Normal $Z_{1j}$} &  &  &  &  \\
$\ $ Oracle & -0.001 & 0.000 & -0.000 & 0.000 \\
$\ $ Ordinary & 1.364 & 1.868 & -0.000 & 0.001 \\
$\ $ Ours ($d_z=1$) & -0.007 & 0.000 & -0.000 & 0.000 \\
$\ $ Ours (ARD) & -0.005 & 0.000 & -0.000 & 0.000 \\
\hline
\textit{Exponential $Z_{1j}$} &  &  &  &  \\
$\ $ Oracle & -0.001 & 0.000 & - & - \\
$\ $ Ordinary & 1.233 & 1.523 & - & - \\
$\ $ Ours ($d_z=1$) & -0.006 & 0.001 & - & - \\
$\ $ Ours ($d_z=2$) & -0.003 & 0.000 & - & - \\
$\ $ Ours (ARD) & -0.005 & 0.000 & - & - \\
\hline
\textit{Binary $Z_{1j}$} &  &  &  &  \\
$\ $ Oracle & -0.000 & 0.000 & - & - \\
$\ $ Ordinary & 1.483 & 2.200 & - & - \\
$\ $ Ours ($d_z=1$) & -0.005 & 0.001 & - & - \\
$\ $ Ours ($d_z=2$) & -0.003 & 0.001 & - & - \\
$\ $ Ours (ARD) & -0.004 & 0.001 & - & - \\
\hline
\textit{Normal $Z_{1j}, Z_{2j}$} &  &  &  &  \\
$\ $ Oracle & -0.001 & 0.000 & 0.000 & 0.000 \\
$\ $ Ordinary & 0.565 & 0.324 & 0.001 & 0.002 \\
$\ $ Ours ($d_z=1$) & -0.111 & 0.013 & 0.001 & 0.000 \\
$\ $ Ours ($d_z=2$) & -0.050 & 0.003 & 0.001 & 0.000 \\
$\ $ Ours ($d_z=3$) & -0.023 & 0.001 & 0.001 & 0.000 \\
$\ $ Ours ($d_z=5$) & -0.008 & 0.001 & 0.001 & 0.000 \\
$\ $ Ours (ARD) & -0.009 & 0.001 & 0.001 & 0.000 \\
\hline
\textit{Normal $Z_{1j}$ and Binary $Z_{2j}$} &  &  &  &  \\
$\ $ Oracle & -0.000 & 0.000 & - & - \\
$\ $ Ordinary & 0.546 & 0.302 & - & - \\
$\ $ Ours ($d_z=1$) & -0.122 & 0.015 & - & - \\
$\ $ Ours ($d_z=2$) & -0.050 & 0.003 & - & - \\
$\ $ Ours ($d_z=3$) & -0.020 & 0.001 & - & - \\
$\ $ Ours ($d_z=5$) & -0.006 & 0.000 & - & - \\
$\ $ Ours (ARD) & -0.007 & 0.001 & - & - \\
\hline
    \end{tabular}
\begin{flushleft}
\footnotesize  NOTE: ATE = average treatment effect; MSE = mean squared error; $d_z$ = the number of latent variables; ARD = latent dimension selected by the ARD prior.
\end{flushleft}
 \end{threeparttable}
\end{table}

\newpage

\section{Unmeasured Cluster-level Moderators}\label{app:simu_mod}

We examine the performance of our VAE-R approach when the unmeasured cluster-level covariate also moderates the treatment effect, without modifying the method. To conceptualize this, we revise our outcome model as follows:
\[
Y_{ij} = m(X_{ij}, W_j) + \delta_Y(Z_j) + \{\tau(X_{ij}, W_j) + \delta_\tau(Z_j)\}\,T_{ij} + \epsilon_{ij} \ ,
\]
where $\delta_\tau(Z_j)$ represent the cluster-level shift in the treatment effect induced by $Z_j$. The treatment model is unchanged from the main text, so that $Z_j$ enters the treatment assignment additively on the logit scale in the confounder setting and not at all in the predictor setting. In this simulation, we generate one normally distributed $Z_{1j} \sim \mathcal{N}(1, 1)$ and set $\delta_\tau(Z_j) = 0.5 Z_{1j}$; all other aspects of the data-generating models follow Section \ref{sec:simu}. 

Since $Z_j$ is unmeasured, we continue to target the CATE defined by measured covariates, which averages over $Z_j$:
\[
\tau^\star(x,w) = E_Z\big[\,E\{Y_{ij}(1) - Y_{ij}(0) \mid X_{ij}=x, W_j=w, Z_j\}\,\big]
= \tau(x,w) + E_Z[\delta_\tau(Z_j)].
\]
Here, $E(Z_{1j})=0.5$, so this reduces to $\tau^\star(x,w) = \tau(x,w)$. The CATE bias and MSE below are computed against $\tau(X_{ij}, W_j)$ as in the main text. 

Table \ref{app:tab:simu_mod} summarizes the simulation results. The oracle uses $Z_j$ in the nuisance models and fits the CATE model on $(X_{ij}, W_j, Z_j)$, then averages the fitted CATE over the empirical distribution of $Z_j$ to target $\tau^\star$; it serves as the performance ceiling. In the confounder setting, the CATE estimates obtained by VAE-R are biased, whereas the AIPW-based ATE estimates show negligible bias owing to the double robustness of the AIPW estimator since the propensity score adjustment is correct. In the predictor setting, the treatment residuals do not depend on $Z_j$, and both the CATE and the AIPW-based ATE estimates show almost no bias, with MSE somewhat larger than in the absence of moderation. 

\begin{table}[H]
    \centering
 \begin{threeparttable}    
    \caption{Simulation results: Moderators \label{app:tab:simu_mod}}
\small
    \begin{tabular}{l|l|C{1.5cm}C{1.5cm}|C{1.5cm}C{1.5cm}}
\hline
\multirow{2}{*}{Design} & \multirow{2}{*}{Condition}  & \multicolumn{2}{c|}{CATE} & \multicolumn{2}{c}{ATE$_{\rm AIPW}$} \\
& & Bias & MSE & Bias & MSE \\
\hline
\multirow{5}{*}{Confounders} & \textit{Normal $Z_{1j}$} &  &  &  &  \\
& $\ $ Oracle & -0.003 & 0.014 & -0.004 & 0.001 \\
& $\ $ Ordinary & 1.494 & 2.405 & 1.533 & 2.361 \\
& $\ $ Ours ($d_z=1$) & -0.112 & 0.044 & -0.007 & 0.001 \\
& $\ $ Ours ($\bar d_{z,ARD}=1.25$) & -0.110 & 0.040 & -0.011 & 0.001 \\
\hline
\multirow{5}{*}{Predictors} & \textit{Normal $Z_{1j}$} &  &  &  &  \\
& $\ $ Oracle & -0.001 & 0.011 & -0.001 & 0.001 \\
& $\ $ Ordinary & -0.001 & 0.073 & -0.001 & 0.002 \\
& $\ $ Ours ($d_z=1$) & -0.001 & 0.019 & -0.001 & 0.001 \\
& $\ $ Ours ($\bar d_{z,ARD}=1.66$) & -0.001 & 0.019 & -0.001 & 0.001 \\
\hline
    \end{tabular}
\begin{flushleft}
\footnotesize  NOTE: MSE = mean squared error; $d_z$ = the number of latent variables; $\bar d_{z,ARD}$ = the average number of latent variables selected by the ARD strategy across 500 replications.
\end{flushleft} 
 \end{threeparttable}
\end{table}

\end{document}